\documentclass{aastex62}
\usepackage{subfigure}
\usepackage{enumitem}
\usepackage{threeparttable}
\usepackage{tabularx}

\received{March 1, 2018}
\revised{May 8, 2018}
\accepted{May 14, 2018}
\submitjournal{AJ}

\shorttitle{Masses and Radii of Very Low Mass Stars}
\shortauthors{Chaturvedi et al.}

\begin{document}

\title{Masses and radii of four very low mass stars in F+M eclipsing binary systems}

\correspondingauthor{Priyanka Chaturvedi}
\email{priyanka.chaturvedi@tifr.res.in}

\author[0000-0002-0786-7307]{Priyanka Chaturvedi}
\affil{Astronomy \& Astrophysics Division, Physical Research Laboratory, Ahmedabad 380009, India}
\affiliation{Department of Astronomy \& Astrophysics, Tata Institute of Fundamental Research, Homi Bhabha Road, Mumbai 400 005, India \\}

\author{Rishikesh Sharma}
\affiliation{Astronomy \& Astrophysics Division, Physical Research Laboratory, Ahmedabad 380009, India}

\author{Abhijit Chakraborty}
\affiliation{Astronomy \& Astrophysics Division, Physical Research Laboratory, Ahmedabad 380009, India}

\author{B.G. Anandarao}
\affiliation{Astronomy \& Astrophysics Division, Physical Research Laboratory, Ahmedabad 380009, India}

\author{Neelam J.S.S.V Prasad}
\affiliation{Astronomy \& Astrophysics Division, Physical Research Laboratory, Ahmedabad 380009, India}

\begin{abstract}
Eclipsing Binaries (EBs) with one of the companions as very low mass stars (VLMS or M dwarfs) are testbeds to substantiate stellar models and evolutionary theories. Here, we present four EB candidates with F type primaries, namely, SAO~106989, HD~24465, EPIC~211682657 and HD~205403, identified from different photometry missions, SuperWasp, KELT, Kepler~2 (K2) and STEREO. Using the high-resolution spectrograph, PARAS, at the 1.2~m telescope at Mount Abu, Rajasthan, India, we hereby report the detection of four VLMS as companions to the four EBs. We performed spectroscopic analysis and found the companion masses to be $0.256\pm0.005$, $0.233\pm0.002$, $0.599\pm0.017$, and $0.406\pm0.005~M_{\odot}$ for SAO~106989, HD~24465, EPIC~211682657 and HD~205403 respectively. We determined orbital periods of $4.39790\pm0.00001$, $7.19635\pm0.00002$, $3.142023\pm0.000003$ and $2.444949\pm0.000001$~d and eccentricities of $0.248\pm0.005$, $0.208\pm0.002$, $0.0097\pm0.0008$ and $0.002\pm0.002$ for EBs SAO~106989, HD~24465, EPIC~211682657 and HD~205403 respectively. The radius derived by modeling the photometry data are $0.326\pm0.012~R_{\odot}$ for SAO~106989, $0.244\pm0.001~R_{\odot}$ for HD~24465, $0.566\pm0.005~R_{\odot}$ for EPIC~211682657, and $0.444\pm0.014~R_{\odot}$ for HD~205403. The radii of HD~24465B and EPIC~211682657B have been measured by precise KEPLER photometry and are consistent with theory within error bars. However, the radii of SAO~106989B and HD~205403B, measured by KELT and STEREO photometry, are $17-20\%$ higher than those predicted by theory. A brief comparison of the results of the current work is made with the M dwarfs already studied in literature. 
\end{abstract}


\section{Introduction} \label{sec:intro}

Studies based on the nature of stellar initial mass function (IMF) have indicated that very low mass stars (VLMS) with masses $\leq$~$0.6~M_{\odot}$ are the most ubiquitous objects created during star formation. IMF was determined for the first time to be a power-law function which decreases with stellar masses in the mass range of $1-10~M_{\odot}$ \citep{Salpeter1955}. Recent work in this field has suggested that stellar IMF breaks from a power-law form at $0.5~M_{\odot}$ with a broad peak between $0.1-0.5~M_{\odot}$ and falling at either side of this mass range \citep{Luhman2000a, Luhman2000b, Kroupa2002, Chabrier2003, Lada2006}. VLMS thus form $\sim~70\%$ fraction of the total stellar systems within a distance of 10 parsec \citep{Henry2006}. With the advent of large infrared arrays, there have been many successful attempts to survey through large field imaging of VLMS objects by many space-based surveys such as Spitzer space telescope mission \citep{Werner2004}, The Two Micron All Sky Survey \citep{Skrutskie2006} (2MASS), Herschel \citep{Pilbratt2010}, and the wide-field infrared survey explorer (WISE) \citep{Wright2010}. These surveys have led to identification and characterization of several VLMS objects \citep{Luhman2012, Gagne2014, Bardalez2015, Gillon2017, Suarez2017, Theissen2017}. Some of the ground-based surveys such as Sloan Digital Sky Survey (SDSS) \citep{York2000}, and The Visible and Infrared Survey Telescope for Astronomy (VISTA) \citep{Emerson2001} too have contributed similarly to the detection of VLMS. However, there have been limited studies on the accurate determination of masses, radii and other physical properties of VLMS due to their instrinsic low luminous nature. 

Testing of various stellar structural and evolutionary models involves precise measurements of physical parameters such as age, mass, radius, temperature, chemical composition of the stars \citep{Torres2010}. Accurate determination of such stellar parameters is possible by studying eclipsing binaries (EBs) by methods such as astrometry, radial velocity (RV) and transit photometry. Such techniques of RV and transit photometry have been applied to hundreds of EBs studied in literature (\cite{Andersen1991, Torres2010} and references therein). A vast majority of observations of M dwarfs for varying masses have reported a higher radius by $10-20\%$ and a lower temperature by $5-10\%$ than those predicted by the models \citep{Chabrier2000, Torres2002, Ribas2003, Lopez-Morales2005, Lopez-Morales2007, Ribas2008, Torres2014, Baraffe2015, Lubin2017}. In particular, the mid M dwarfs (M3) that form the boundary between stars having radiative zone and those with totally convective zones \citep{Chabrier2000} are reported to show the most glaring discrepancies in the measurement of radii when compared with the theoretical models \citep{Lopez-Morales2007}. The mismatch of the radii as seen in these stars is termed as the `M dwarf radius problem' \citep{Triaud2013}. 

Stellar activity hypothesis suggests that the afore-mentioned disagreement between theory and observations may be primarily caused by the degree of magnetic activity in stars: strong magnetic fields inhibit convection leading to inflated stellar radii (\citealt{Lopez-Morales2005}; \citealt{Mullan2001}; \citealt{Torres2013}). \cite{Lopez-Morales2005} propose a scenario based on energy conservation mechanisms in star spot-covered areas. The dynamo-generated magnetic fields affect the convectional stability criteria for the stars leading to a bloated radius at the same temperature or lower temperatures for the same radius. There is an inherent assumption that strong magnetic field regions and star spots are cooler than their surroundings. Thus, the suppressed photospheric temperatures lead to measured inflated stellar radii in order to maintain the radiative equilibrium and hydrostatic equilibrium. \cite{Chabrier2007} concluded in their study that the inhibition of convection in fast rotating stars and the presence of star spots on the stellar disk could affect the stellar models. Current atmospheric models are not accurate due to some missing opacity components leading to larger radius for stars having higher metallicity. \cite{Berger2006} in their study find that the disagreement is larger among metal-rich stars than metal-poor stars. This hypothesis suggests the dependency of metallicity on the amount of inflation for the measured radius.

Double-lined EBs, specifically M-M EBs having masses and radii determined at high accuracies ($\sim2\%$) like CM~Dra \citep{Morales2009}, Cu~CnC \citep{Ribas2003}, YY~Gem \citep{Torres2002}, Gu~Boo \citep{Lopez-Morales2005} are paradigms used to test observations against theoretical models. The fundamental parameters of M dwarfs have been determined by variety of methods including spectral energy distribution, a combination of photometric and spectroscopic parameters and similar such methods. Comparing the best studied M dwarfs in EBs with the theoretical models using a range of isochrones of different ages and metallicities have seemed to reduce the scatter seen in the mass-radius diagram of M dwarfs \citep{Torres2013}. In order to further reduce the scatter, there is a need to have stellar parameters derived with high accuracies and precision for a range of systems by different methods. Single-lined detached EB systems where VLMS objects occur as companions to brighter F, G and K type stars, provide a huge sample to fill the gap from observations. RV and transit photometry techniques ensure indirect determination of stellar parameters at high accuracies. RV technique applied on F, G, K type primaries help determine the projected mass of the companion whereas photometry of these targets gives insights on the angle of orbital inclination and radii of both the components providing us a complete picture of the EB system. F-type primaries accompanied with M-type secondaries (hereafter F+M binaries) in EBs are very often discovered in photometric surveys as they have resemblance to hot Jupiters transiting main sequence stars (e.g. \cite{Bouchy2005}; \cite{Beatty2007}). However, only a handful of F$+$M EBs have been studied for their masses, radii and orbital parameters (e.g. \cite{Pont2005a, Pont2005b, Pont2006, Fernandez2009, Chaturvedi2014}). Statistically there is a higher probability of finding M dwarfs in companion with F-type primaries in contrast to finding equal mass binary pairs \citep{Moe2015, Bouchy2011, Bouchy2011b}. In order to understand the binarity fraction for F and M type stars, every additional system discovered and analyzed plays a key role in making the sample of F$+$M binaries larger and thereby an important subset of stellar studies. \cite{Duquennoy1991} found a $57\%$ binarity fraction for late F and G type stars for a distance limited sample within 22 pc of the Sun. Similar studies on F type stars ($1.1~M_{\odot}~\geq~M~\leq~1.7~M_{\odot}$) by \cite{Fuhrmann2012, Fuhrmann2015} have found that majority of F type stars ($\sim~2/3$ of them) are multiple by nature. F type stars have a range of rotational velocities \citep{Nordstrom2004} and the exteriors of these stars range from having convective envelopes (late F type stars) to radiative envelopes (mid to early F type stars). Stars massive than F type (O, B and A type) are difficult to study for their binarity because of their high stellar rotation rates and relatively smaller sample size. The higher temperatures lead to less number of photospheric lines in the spectra. Moreover, these lines are rotationally broadened thereby decreasing the quality of stellar spectra and decreasing spectroscopic precision. Many of the early type stars have stellar pulsations making detection of companions very difficult. There have been only handful of such detections, for example the case of WASP-33b \citep{Herrero2011} and references therein). Thus F type stars occurring in binary pairs will be valuable contributions for understanding the multiplicity fraction of early type stars.

We have initiated the EB program by PRL Advanced Radial velocity Abu-sky Search (PARAS) \citep{Chakraborty2014} with a motivation to study single-lined detached EBs having potential M dwarf companions. We have previously reported the detection and characterization of two EBs with PARAS, a $0.286~M_{\odot}$ M dwarf across a F-type primary \citep{Chaturvedi2014} and a $0.098~M_{\odot}$ late-type M dwarf across a K-type primary \citep{Chaturvedi2016}. In this third paper of the series, we present spectroscopic and photometric investigations on four F-type sources, SAO~106989, HD~24465, EPIC~211682657, and HD~205403. All these stars are EBs of short orbital period with putative M dwarfs in orbit. This study is intended to determine the masses, radii and orbital parameters of the four putative M dwarfs. We describe the program stars briefly in \S~2. This is followed by a description of RV observations of stars in \S~3. We also discuss the high-resolution spectroscopic and photometric methods of analysis used to derive the physical parameters concerning all the EBs in this section. In \S~4 we discuss the importance of this work followed by a brief summary in \S~5.

\section{Program Stars and Observations} \label{sec:program}

Selection of EB candidates involved choosing stars brighter than $\sim$~11 in the V-band, as it is the faintest limit for PARAS with the 1.2~m telescope. Spectral types from F to K type were chosen as these stars have more spectral lines for precise RV measurements. Candidates were also chosen based on its coordinates in the non-monsoon months between October-May of the observing season at Mt. Abu. The current interest being VLMS, candidates having an upper cut-off for the transit depth at $\sim$~50~mmag have been chosen in order to avoid samples having massive secondaries as companions. The lower limit cut-off of the transit depth while shortlisting candidates is kept at $\sim$ 12~mmag to avoid planetary candidates. Based on these selection criteria, nearly a dozen targets have been shortlisted from a list of few hundreds of EB candidates picked up from various photometric surveys like \textit{STEREO} \citep{Wraight2012}, \textit{SuperWASP} \citep{Street2007, Christian2006, Lister2007, Clarkson2007, Kane2008}, and \textit{Kepler} \citep{Barros2016}.

\textit{STEREO}, Solar Terrestrial Relations Observatory, are two spacecrafts (A $\&$ B) primarily dedicated to look at Sun and it's environment. The Heliospheric Imager (HI-1) on the Ahead spacecraft (HI-1A) has been used to study variability of stars up to 12 mag \citep{Wraight2011}. About 263 EB candidates have been made public after a survey of 650,000 stars with magnitudes brighter than 11.5. \textit{SuperWASP} (SW) is an extra-solar planet detection programme hosted by the joint collaboration between eight academic institutes located in the United Kingdom~\footnote{www.superwasp.org}. SW consists of ground-based robotic observatories and eight wide-angle cameras covering both the hemispheres of the sky. SW-N is located on the island of La Palma among the Isaac Newton Group of telescopes (ING) and SW-S at the site of the South African Astronomical Observatory (SAAO). The operational wavelength band is the entire V band covering stars having magnitudes between 8 to 15 listing several exoplanet candidates; few of which have been speculated by us as potential EB hosts. \textit{Kepler} is a space observatory launched by NASA on March 7, 2009 to discover Earth-like planets orbiting other stars. Along with the usual target list of potential exoplanet host stars, \textit{Kepler} also published a catalogue of eclipsing binary candidates \citep{Koch2007}. The mission is designed specifically to look at around 100,000 stars for transits in the region above the galactic plane looking down at the Orion arm of the Milky Way galaxy \citep{Borucki2009}. The aim was to look at the sources which have been flagged as `EB' in the \textit{Kepler} catalogue. 

One of the four program stars chosen for the study, \textbf{SAO~106989}, is shortlisted from the ground-based SuperWasp (SW) photometry catalogue \citep{Street2007}. This source was an exoplanet candidate having a periodicity of 4.4~d and a transit depth of 13.5~mmag. Based on the radius estimation, the secondary was speculated to be a hot Jupiter or a M dwarf companion (both the objects have comparable sizes). \textbf{HD~24465} and \textbf{EPIC~211682657} are shortlisted from K2 photometry database. These candidates are reported to have periodicities of 7.19 and 3.142~d and transit depths of 38 and 46 mmag respectively \citep{Barros2016}. For HD~24465 and EPIC~211682657, K2 data are available. The periodicity for \textbf{HD~205403}, which is shortlisted from STEREO catalog, is 2.44~d with a transit depth of 57 mmag \citep{Wraight2012}. The stellar parameters for all the sources from previous studies are listed in Table~\ref{tab:result_sao}.

\subsection{RV observations}
\label{sec:rv_obs}

High-resolution spectroscopic observations of the program EBs were taken during 2013-2017 using the optical fiber-fed echelle spectrograph, PARAS, (high resolution, R~$\sim$~67,000, cross-dispersed spectrograph) coupled with the 1.2~m telescope at Gurushikhar, Mount Abu, India. The spectrograph has a spectral coverage of $3800-9000~\AA$. However, for precise RV measurements, wavelength range of $3800-6800~\AA$~is utilized. The spectra are recorded in the simultaneous reference mode, wherein one of the two optical fibers are illuminated by the target source and the other is illuminated with Thorium-Argon (ThAr) as the calibration lamp. The spectrograph is maintained in a temperature-stable (RMS of 0.01$^\circ$C at 25$^\circ$C) and pressure-stable environment (maximum variation of 0.06 mbar in one night of observation). The nightly calibration sequence includes 5 bias frames and 3 flat frames (for which both fibers are illuminated with a tungsten lamp), and several ThAr-ThAr frames (for which both fibers are illuminated with the calibration lamp) throughout the night. The purpose of the ThAr-ThAr frames is to carefully measure absolute instrument drift, as well as differential drifts. Science observations are usually made using simultaneous star-ThAr exposures ($2-3$ exposures per night per target). Details of the spectrograph, observational procedure, and data analysis techniques can be found in \cite{Chakraborty2014}.

A total of 17 sets of observations of the source SAO~106989 were acquired between October to November 2013 at a resolving power of 67000. During 2013, due to telescope tracking issues, it was difficult to keep exposure durations more than 1200~s despite of closed cycle on-axis star guiding. This resulted in SNR between 12 to 22 per pixel at the blaze peak wavelength of the spectrum at 550 nm. This problem of telescope tracking was solved in late 2014 and data taken post that were free from such issues. The source was observed on all nights at an air mass between 1.1-1.3. For star HD~24465, 14 sets of observations were acquired between October to December 2016. Based on sky conditions, the exposure duration on most of the nights was 3000~s whereas for some nights it was kept at 1800~s resulting in SNR ranging between 13 to 30 per pixel at the blaze peak wavelength of the spectrum at 550 nm. The air mass throughout the observations for this source was between 1.01-1.55. For the source, EPIC~211682657, 18 spectra were recorded between the months of May and November 2017. The SNR for these spectra were between 21-30 per pixel at the blaze wavelength around 550 nm with an exposure time of 2400~s. The air mass for the observations on this particular source was between 1.1-1.2. In a similar way, HD~205403, was also observed between the months of May and November 2017. The SNR per pixel at the blaze wavelength around 550 nm for each exposure was between 31-40 depending on the exposure times ranging between 1800-2400~s. The air mass varied between 1.5-1.6 during the course of observations for HD~205403. All nights of observations were spectroscopic in nature with cloud cover less than $\sim~40\%$ and nightly seeing less than or equal to 2.0 arcsec. A list of epochs and observational details for all the stars are shown in Table~\ref{tab:rv_sao}. The first two columns represent the observation time (mid-exposure) in UT and BJD respectively. The exposure time and observed RV are given in the following columns. The RV errors are limited by photon noise is as given by \cite{Hatzes1992} $\sigma_{\text{RV}} \sim 1.45\times10^{9} (S/N)^{-1} R^{-1} B^{-1/2} \text{ m s}^{-1}$. Here, $S/N$ is the signal-to-noise of the spectra, while $R$ and $B$ are the resolving power and wavelength coverage of the spectrograph in angstrom~(\AA) respectively. In order to compute errors on RV, we randomly varied the signal on each pixel within the Poissonian uncertainty of $\pm\sqrt{N}$, where N is the signal on each pixel, and thereafter computed the CCF for each spectra. This process is repeated 100 times for each spectra and the standard deviation of the distribution of the obtained RV values is given as the 1~$\sigma$ uncertainty on the CCF fitting along with errors from photon noise on each RV point. The computed RV errors are given in the last column of Table~\ref{tab:rv_sao}. 

\subsection{Photometry observations}

In order to determine the radii of both the components of the EBs, transit photometry is a suitable technique. All the stars have been observed by ground-based or space-based photometry missions previously. SAO~106989 was first listed as an exoplanet candidate from SuperWasp (SW) photometry catalogs after surveying millions of stars in the night sky \citep{Street2007}. While the SW photometry has listed many exoplanet candidates in short periods between $2-3$ days \citep{Street2007}, a periodicity of 4.4~d is relatively long for the catalog's sampling standards. Thereby, due to inadequate time cadence, less number of transit datapoints is recorded for this source. Moreover, the data for this source looks noisy as seen in \citep{Street2007}.  However, we found the source had been observed by The Kilodegree Extremely Little Telescope (KELT) survey \citep{Pepper2007} \footnote{http://exoplanetarchive.ipac.caltech.edu}. KELT consists of two robotic telescopes for conducting a survey for transiting exoplanets around bright stars. The telescope is a wide-field ($26~\times~26$ square degress), small aperture (42.0~mm) system optimized for imaging bright stars in a broad R-band. The telescope is not tracking any field and thus a single field is imaged for 1-2~h with an average precision of 7.5~mmag on each observing night. We analyzed the z-band image from SDSS-III~\footnote{http://skyserver.sdss.org/} to check for a possibility of third light contamination. The SDSS-III images are $6~\arcmin~\times~10~\arcmin$ and have a plate scale of $0.4~\arcsec/pixel$ \citep{Gunn1998}. The image is not centred at the source. We have marked circles of radii 30~\arcsec, 60~\arcsec and 120~\arcsec from the centre of the source for identifying possible contaminants. There is no contaminant seen within 1~\arcmin. The nearest source resolved within $2~\arcmin$ is TYC 1658-738-1 and has a magnitude difference of $\Delta~V = 2.21$ from our source of interest, SAO~106989. The plate scale of KELT is $23~\arcsec/pixel$ and a photometric aperture of $3\arcmin$ \citep{Siverd2012}, a source between 1--2$\arcmin$ could cause slight light contamination making the photometry data appear noisy. However, there is no evidence of photometric dip as predicted during the time of secondary eclipse. This rules out a light blending scenario but we would warn the readers of light contamination.

\begin{figure*}[!ht]
\vspace{-4.5cm}
\centering
	\includegraphics[width=0.5\textwidth]{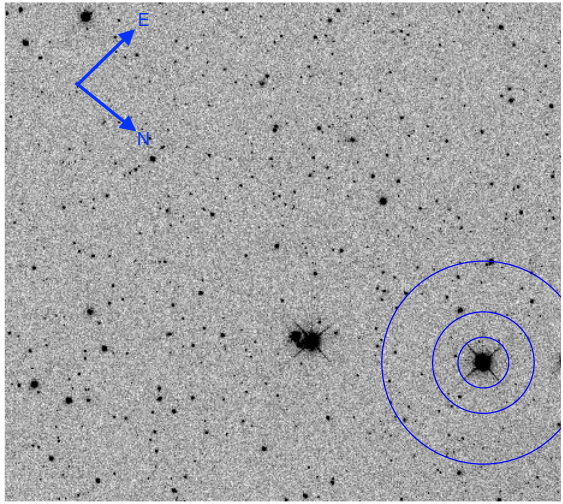}
\caption{SDSS z-band image for the source SAO~106989. The radius of inner, middle and outer blue circle centred on the source are $30~\arcsec$, $1~\arcmin$ and $2~\arcmin$, respectively. It can be seen that there is no potential source of contamination within 30~\arcsec. The nearest possible source of contamination ($\Delta~V = 2.21$ mag) is between $1~\arcmin$ and $2~\arcmin$. \label{fig:dss_image}}
\end{figure*}

HD~24465 and EPIC~211682657 are K2 candidates. Kepler mission launched in 2009 \citep{Borucki2009} has led to a surge in the detection of exoplanets and EB candidates. The K2 mission, being a successor of the former Kepler mission, the number of detections have grown exponentially, as K2 observes 4 fields in one year and the targets observed are on an average brighter than Kepler candidates. The photometry data is taken in Kepler filter with a wavelength range between 4200-8900~\AA ($\lambda_c = 6400~\AA$) \citep{Brown2011} timed between 08 February -- 20 April 2015 for HD~24465 and 27 April -- 10 July 2015 for EPIC~211682657 with an average photometry precision of 15 ppm. HD~205403 was one of the nine candidates shortlisted from the NASA STEREO mission as a part of the bright eclipsing candidates (visual magnitude 6~$<$~V~$<$~12) surveyed by the two satellites onboard the STEREO mission looking for stars with effective temperatures between 4000 and 7000 K \citep{Wraight2012}. This program star is an EB candidate, which has a companion radius predicted to be greater than $0.35$~R$_{\odot}$. The star was observed in the wavelength band between 630-730~nm with an exposure duration of 40~s. The data was taken every 40~m for the complete duration of 16 to 17~d when the star was observable on the CCD FOV for each cycle of observation (roughly a year).

\section{Data Modeling and Results} \label{sec:results}
In this section, we describe the analysis techniques used to reduce the data and the methodology utilized to determine the orbital parameters of the stars studied in this paper.

\subsection{Radial Velocity of the primary stars of the EB systems}

%------------------------------------------------------------------------------------------------
%  FIG 1
%------------------------------------------------------------------------------------------------
\begin{figure*}[!htbp]
       \includegraphics[width=0.5\textwidth]{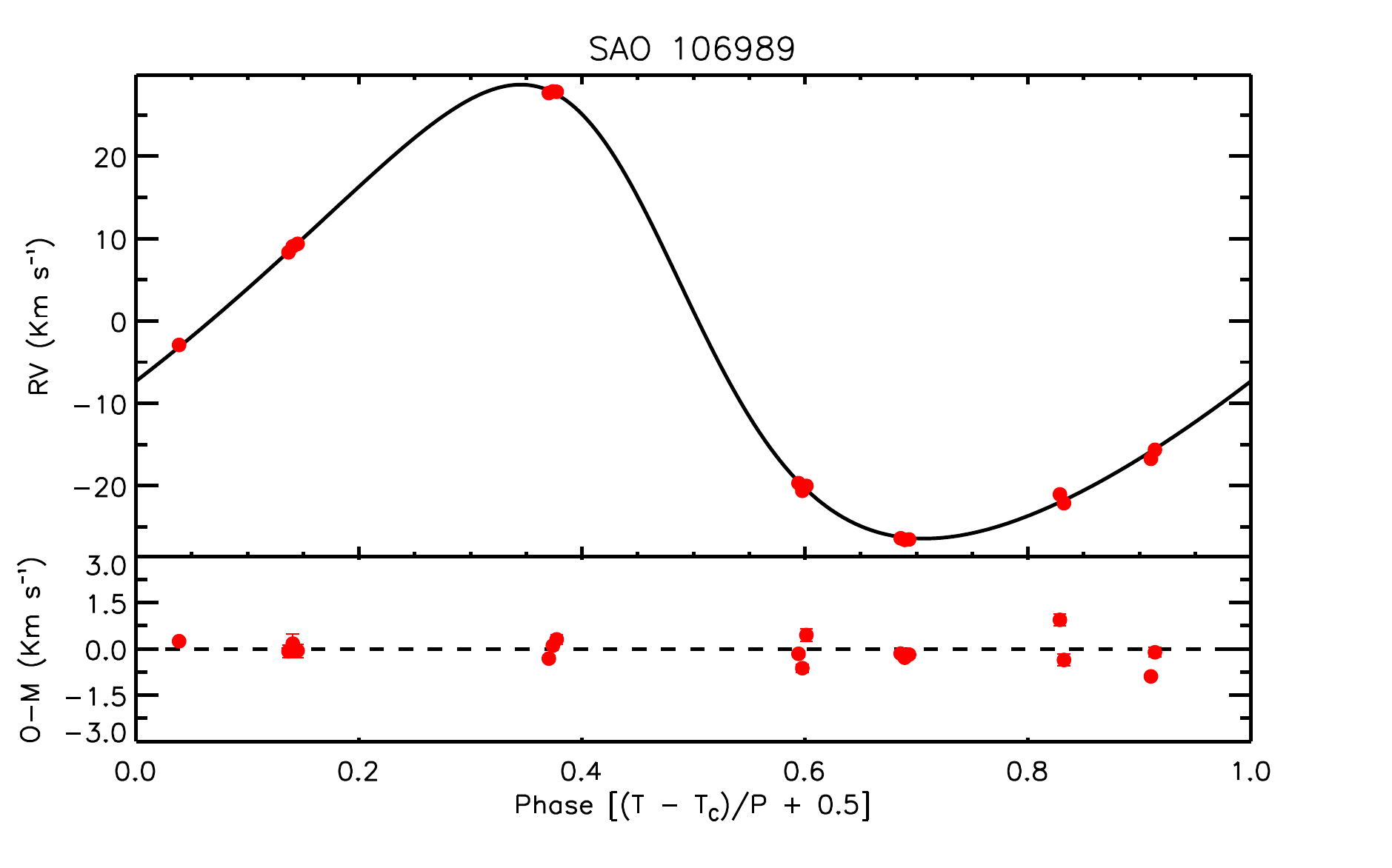}
	\includegraphics[width=0.5\textwidth]{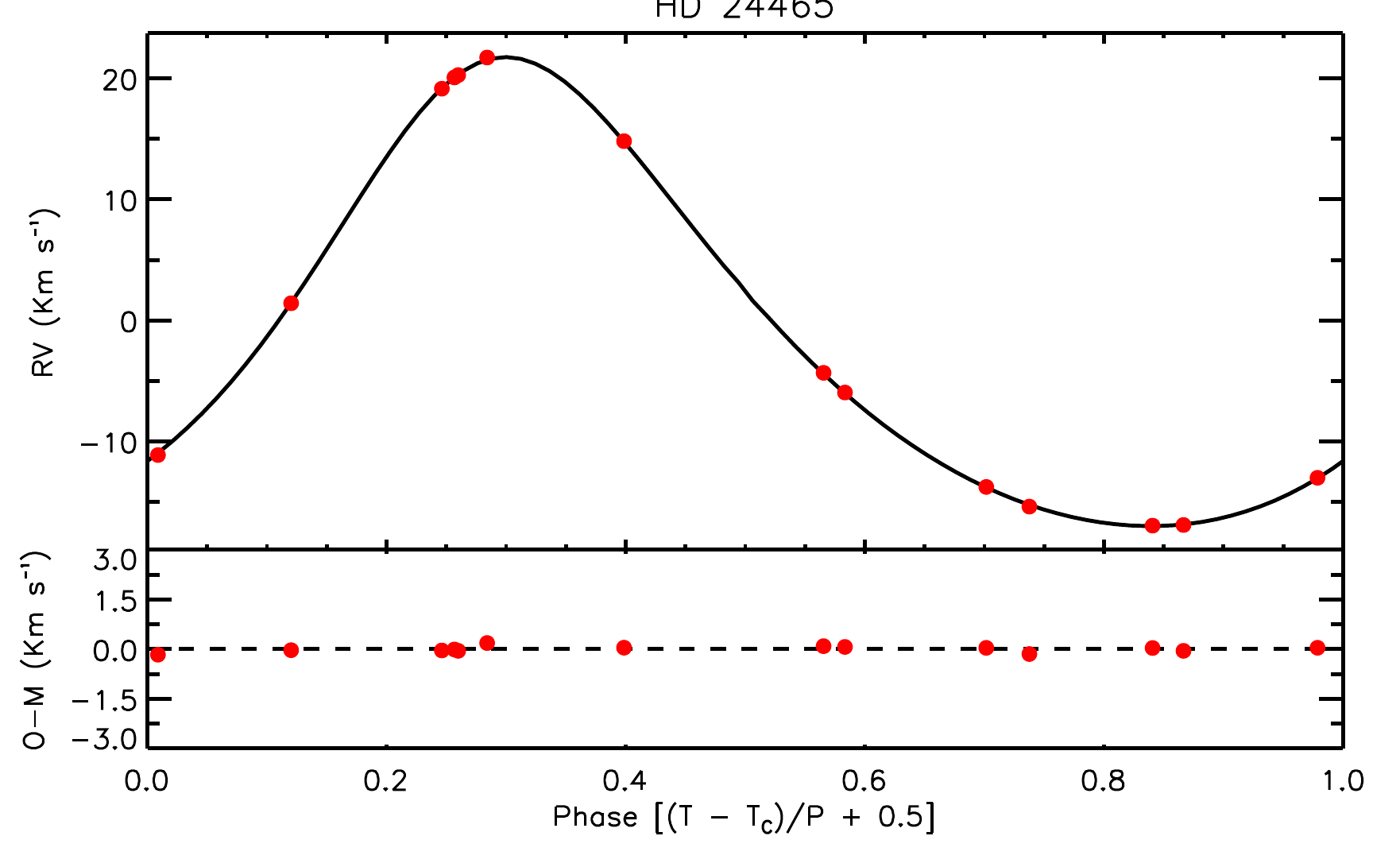}
	\includegraphics[width=0.5\textwidth]{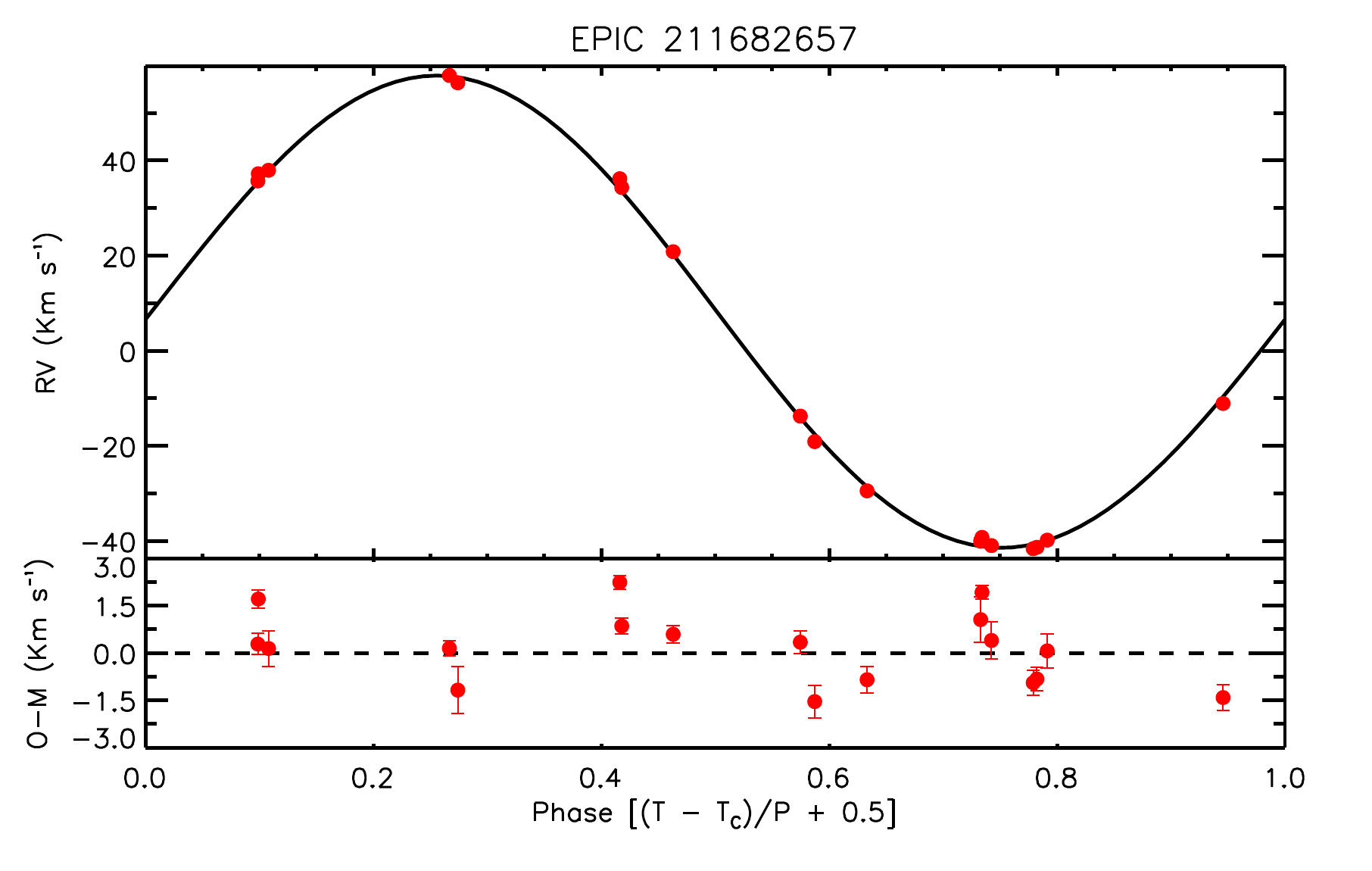} 
	\includegraphics[width=0.5\textwidth]{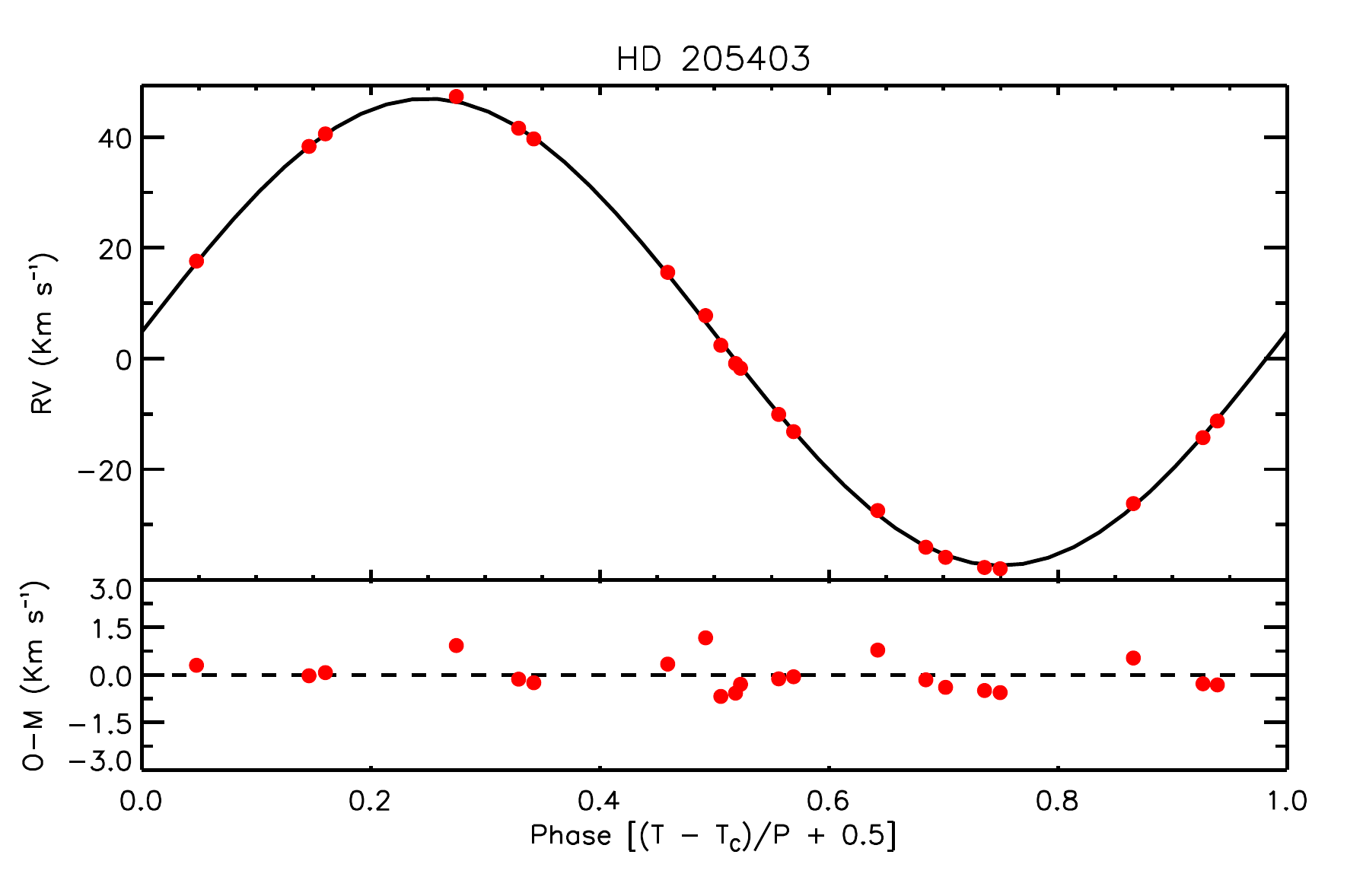}
\caption{(Top panel of each subfigure) RV model curve for star SAO~106989 (top left), HD~24465 (top right), EPIC~211682657 (bottom left) and HD~205403 (bottom right) obtained from PHOEBE (Refer \S~\ref{subsec:orbital_sao} for details on PHOEBE) is plotted against orbital phase. PARAS, Mount Abu (solid red circles) observed data points along with the estimated errors are overplotted on the curve.
(Bottom panel of each subfigure ) The residuals from best-fitting are plotted below the RV plot. For better visual representation, the x axis in Phase is shifted by 0.5 so that the central primary transit crossing point (T$_c$) occurs at phase 0.5 instead of 0. \label{fig:rv_sao}}
\end{figure*}

%------------------------------------------------------------------------------------------------

The barycentric corrected RV values and their respective uncertainties are shown in Table~\ref{tab:rv_sao} for all the stars studied in this paper. The phase-folded RV points for all the stars are plotted in Fig~\ref{fig:rv_sao}. The figure shows four panels; starting from SAO~106989 in top left, HD~24465 in top right, EPIC~211682657 in bottom left and HD~205403 in bottom right. The red circles in each panel are the observed RV points and the solid line is the fitted model for each star, details of which are discussed in \S~\ref{subsec:orbital_sao}. Based on temperatures determined from \S~\ref{sec:spec_analysis}, SAO~106989 and HD~24465 are found to be F9/G0 and F7/F8 type stars respectively \citep{Pecaut2013}. Thus, a G2 numerical mask was used as the zero velocity cross-correlation template to compute RV measurements. The other two stars, EPIC~211682657 and HD~205403 are found to be F4 and F5 type respectively based on their temperatures \citep{Pecaut2013} and hence F5 numerical mask was used for the cross-correlation.

The data extraction and analysis pipeline (PARAS PIPELINE) is a set of routines written in {\tt{IDL}} to ease the complex and time consuming process of data reduction. It is fully automated requiring minimal amount of user interaction; only if external factors necessitate it. PARAS PIPELINE is based on the {\tt{REDUCE}} data analysis package developed by \cite{Piskunov2002} for processing cross-dispersed echelle data. It is modified to suit the requirements of PARAS data. The reduction process requires intake of bias, flat fields and calibration lamp frames in unison with the science exposures. Bias frames are for bias corrections whereas flat frames are for the purpose of order location taken by illuminating the fiber by hot Tungsten lamp generally before the science exposures. Wavelength calibration is accomplished by comparing the observed arc lamp (ThAr for current case) spectrum with a suitable template spectrum. The wavelength solution was generated for simultaneously illuminated ThAr lamp spectra and can be used as a blueprint solution as long as external modifications do not affect the fiber and its position. A complete thorium line list for the PARAS spectral range is utilized (similar to the SOPHIE line list at www.obs‑hp.fr). For automated process, a binary mask of sharp thorium lines is created which is used to assist the calibration process. The cross-correlation function (CCF) is calculated by shifting this thorium mask against each spectral order, and the net drift value is corrected for each spectra. The extracted wavelength solution is imposed on the observed stellar spectra in the simultaneous reference mode, thereby enabling wavelength solution for each observed science exposure by incorporating necessary drift corrections. RVs are finally derived by cross-correlating target spectra, i.e. computing the CCF with a suitable numerical stellar template mask, created especially from high signal-to-noise ratio spectra or synthetic data \citep{Baranne1996}. It consists of values 1 and 0, where non-zero values correspond to theoretical positions and widths of the absorption lines at zero velocity. The CCF is constructed by shifting the mask as a function of Doppler velocity. RVs are then corrected for their barycentric velocities. For complete details on the reduction and analysis methods, readers are requested to follow \citep{Chakraborty2014}.

%------------------------------------------------------------------------------------------------
%  TABLE 1
%------------------------------------------------------------------------------------------------

\startlongtable
\begin{deluxetable}{lcccc||clcccc}
\tiny
\tabcolsep=0.11cm
\tablewidth{0.1pt}
\tablecaption{RV Observations for all the stars : SAO~106989, HD~24465, EPIC~211682657 and HD~205403 \label{tab:rv_sao}}
\tablehead{
\colhead{UT Date} & \colhead{T-2,400,000} & \colhead{Exp. Time} & \colhead{RV} & \colhead{$\sigma$-RV} & & \colhead{UT Date} & \colhead{T-2,400,000} & \colhead{Exp. Time} & \colhead{RV} & \colhead{$\sigma$-RV}\\
\colhead{} & \colhead{(BJD-TDB)} & \colhead{(sec.)} & \colhead{(km s$^{-1}$)} & \colhead{(km s$^{-1}$)} & & \colhead{} & \colhead{(BJD-TDB)} & \colhead{(sec.)} & \colhead{(km s$^{-1}$)} & \colhead{(km s$^{-1}$)}\\
}
\startdata
\textbf{SAO~106989} &&&&  & & \textbf{EPIC~211682657} &&&& \\
2013 Oct  22 & 56588.193  & 1200	& $-$16.373    & 0.116	 && 2017 May 04  &  57878.130  & 2400  &   $-$21.050	&  0.393   \\
2013 Oct  22 & 56588.209  & 1200	& $-$16.566   & 0.086   && 2017 May 04  &  57878.169  & 2400  &  $-$18.993  &  0.535    \\
2013 Oct  22 & 56588.226  & 1200 	& $-$16.517    & 0.099   && 2017 May 05  &  57879.135  & 2400  &    56.378  &  0.333   \\
2013 Oct  23 & 56589.180  & 1200 	& $-$6.733    & 0.107   && 2017 May 05  &  57879.164  & 2400  &    58.536	&  0.563    \\
2013 Oct  23 & 56589.196  & 1200	& $-$5.638    & 0.151   && 2017 May 06  &  57880.132  & 2400  &    53.411  &  0.485   \\
2013 Oct  24 & 56590.176  & 1200	& 18.325        & 0.202   && 2017 May 07  &  57881.128  & 2400  &   $-$19.283	&  0.723	\\
2013 Oct  24 & 56590.193  & 1200	& 19.055        & 0.318   && 2017 May 07  &   57881.158 &  2400 &    $-$20.279 &  0.606    \\
2013 Oct  24 & 56590.212  & 1200 	& 19.362      & 0.215 	 && 2017 Oct 23  &   58050.483 &  2400 &   $-$8.728  & 0.418 \\
2013 Oct  25 & 56591.203  & 1200	& 37.663      & 0.102 	 && 2017 Oct 24  &   58051.465 & 2400  &    9.638    &  0.396 \\
2013 Oct  25 & 56591.219  & 1200	& 37.857       & 0.113   && 2017 Oct 25  &   58052.474 & 2400  & 78.608      &  0.238 \\
2013 Oct  25 & 56591.234  & 1200	& 37.813      & 0.160   && 2017 Oct 25  &   58052.497 & 1200  & 77.067     & 0.743 \\
2013 Oct  26 & 56592.188  & 1200	& $-$9.680    & 0.098   && 2017 Oct 26  &   58053.441  & 2400  & 6.985  & 0.362 \\
2013 Oct  26 & 56592.203  & 1200	& $-$10.613    & 0.133   && 2017 Oct 26 &    58053.481   & 2400 & 1.624   &  0.522 \\
2013 Oct  26 & 56592.219  & 1200 	& $-$10.016    & 0.214   && 2017 Nov 23 &     58081.369  & 2400 & 41.554  & 0.273 \\
2013 Oct  27 & 56593.219  & 1200 	& $-$11.047   & 0.191   && 2017 Nov 24  &    58082.372  &  2400 & $-$20.593 & 0.368 \\
2013 Oct  27 & 56593.235  & 1200 	& $-$12.113    & 0.192   && 2017 Nov 25 &     58083.867  & 2400 & 57.949 & 0.285 \\
2013 Nov  19 & 56616.132  & 1200 	&  7.089    & 0.052   && 2017 Nov 26  &   58084.369 & 2400 & 55.053 & 0.255 \\
&&&&&& 2017 Nov 27  &   58085.363 & 2400 & $-$18.512 & 0.211 \\
\hline
\\\textbf{HD~24465} &&&& & & \textbf{HD~205403} &&&& \\
2016 Oct  20 & 57682.473  & 3000	& $-$29.703    & 0.047	 && 2017 May 03  &  57877.475 &  2400  &   $-$17.531 &    0.118  \\  
2016 Oct  21 & 57683.400  & 3000	& $-$31.777    & 0.022   && 2017 May 04  &  57878.466 &  2400  &    27.566 	&   0.035   \\ 
2016 Oct  22 & 57684.425  & 3000 	& $-$26.080    & 0.027   && 2017 May 05  &  57879.471 &  1800  &    25.545  &   0.040   \\
2016 Oct  24 & 57686.406  & 3000 	& 6.609    	& 0.033   && 2017 May 06  &  57880.466 &  1800  &   $-$16.253 &   0.101  \\ 
2016 Nov  24 & 57717.345  & 1800	& $-$20.181    & 0.017   && 2017 May 07  &  57881.465 &  2400  &    57.207  &    0.070  \\
2016 Dec  01 & 57724.412  & 3000	&$-$ 19.660     & 0.015   && 2017 Oct 24  &  58051.169 &  2400  &    $-$24.082 &  0.054 \\
2016 Dec  02 & 57725.391  & 3000	& $-$29.201     & 0.011  && 2017 Oct 24  &  58051.211  & 2400  &    $-$25.807  & 0.063 \\
2016 Dec  03 & 57726.392  & 3000	& $-$32.479     & 0.014   && 2017 Oct 26  &  58053.143  & 2400  &    17.776 &   0.063  \\
2016 Dec  04 & 57727.385  & 3000	& $-$28.576     & 0.013   && 2017 Oct 26  &  58053.176  & 2400  &    12.327  &   0.063  \\
2016 Dec  05 & 57728.404  & 1800	& $-$14.190     & 0.012   && 2017 Oct 26  &  58053.207  & 2400  &    9.18    &   0.073 \\
2016 Dec  06 & 57729.311  & 1800	& 3.471    & 0.015   && 2017 Nov 22  &  58080.113  &  2400 &    8.2811  &   0.052 \\
2016 Dec  06 & 57729.387   & 1800	& 4.411    & 0.016   && 2017 Nov 23  &  58081.100  &  2400 &    $-$4.205 &  0.053  \\
2016 Dec  06 & 57729.410   & 1800	& 4.624    & 0.015   && 2017 Nov 23  &  58081.130  &  2400 &    $-$1.091  & 0.063 \\
2016 Dec  07 & 57730.408  & 1800	& $-$0.710    & 0.015   && 2017 Nov 24  &  58082.084  &  2400 &    51.576    & 0.052 \\
&&&&&& 2017 Nov 24  &  58082.115  & 2400  &    49.116   &   0.052 \\
&&&&&& 2017 Nov 25  & 58083.078 & 2400  &      $-$27.789 & 0.053 \\
&&&&&& 2017 Nov 25  & 58053.112 & 2400  & $-$28.016 & 0.053 \\
&&&&&& 2017 Nov 26  & 58084.081  & 2400 & 48.109  & 0.063 \\
&&&&&& 2017 Nov 26  & 58084.116  & 2400  & 50.53   & 0.073 \\
&&&&&& 2017 Nov 27  & 58085.084  &  2400 & 0.055  &  0.063 \\
&&&&&& 2017 Nov 27  & 58085.116  & 2400  &  $-$3.132 & 0.073 \\
\hline
\enddata
\end{deluxetable}
%--------------------------------------

\subsection{Spectral synthesis} \label{sec:spec_analysis}

We utilized the stellar synthesis pipeline, \textit{PARAS SPEC}, to estimate the stellar parameters from the observed spectra \citep{Chaturvedi2016}. Spectra obtained from the instrument are unblazed by fitting a polynomial function and stitched across the echelle orders to produce a single spectra. Individual spectra are normalized and many epochs of the same star are co-added after relevant RV corrections to get a high S/N continuum normalized stitched stellar spectra. This spectra serves as input to the stellar pipeline. The observed spectra needs to be compared against a grid of synthetic spectra. This grid is produced using synthetic spectra generator code {\tt{SPECTRUM}}. It utilizes the ATLAS9 models by Kurucz \citep{Kurucz1993} for stellar atmosphere parameters by working on the principle of local thermodynamic equilibrium and plane parallel atmospheres. The library consists of synthetic spectra having $T_{\rm{eff}}$ between $4000-7000$ K at an interval of $250$ K, $[Fe/H]$ in a range of $-2.5 - 0.5$ dex with an interval of $0.5$ dex and ${\rm log}~g$ between $1.0-5.0$ dex with an interval of $0.5$ dex. The wavelength range for the generated synthetic spectra is kept between $5050-6560$ with an interval of $0.01$~\AA~and a velocity resolution of 1~km s$^{-1}$ between $1-40$~km s$^{-1}$. \textit{PARAS SPEC} is based on two primary methods, synthetic spectral fitting method and equivalent width method. These two methods and the results obtained after applying these methods on the target sources are briefly discussed here.

%------------------------------------------------------------------------------------------------
%  FIG 2 and FIG 3
%------------------------------------------------------------------------------------------------

\begin{figure*}[!htbp]
\includegraphics[width=0.48\textwidth]{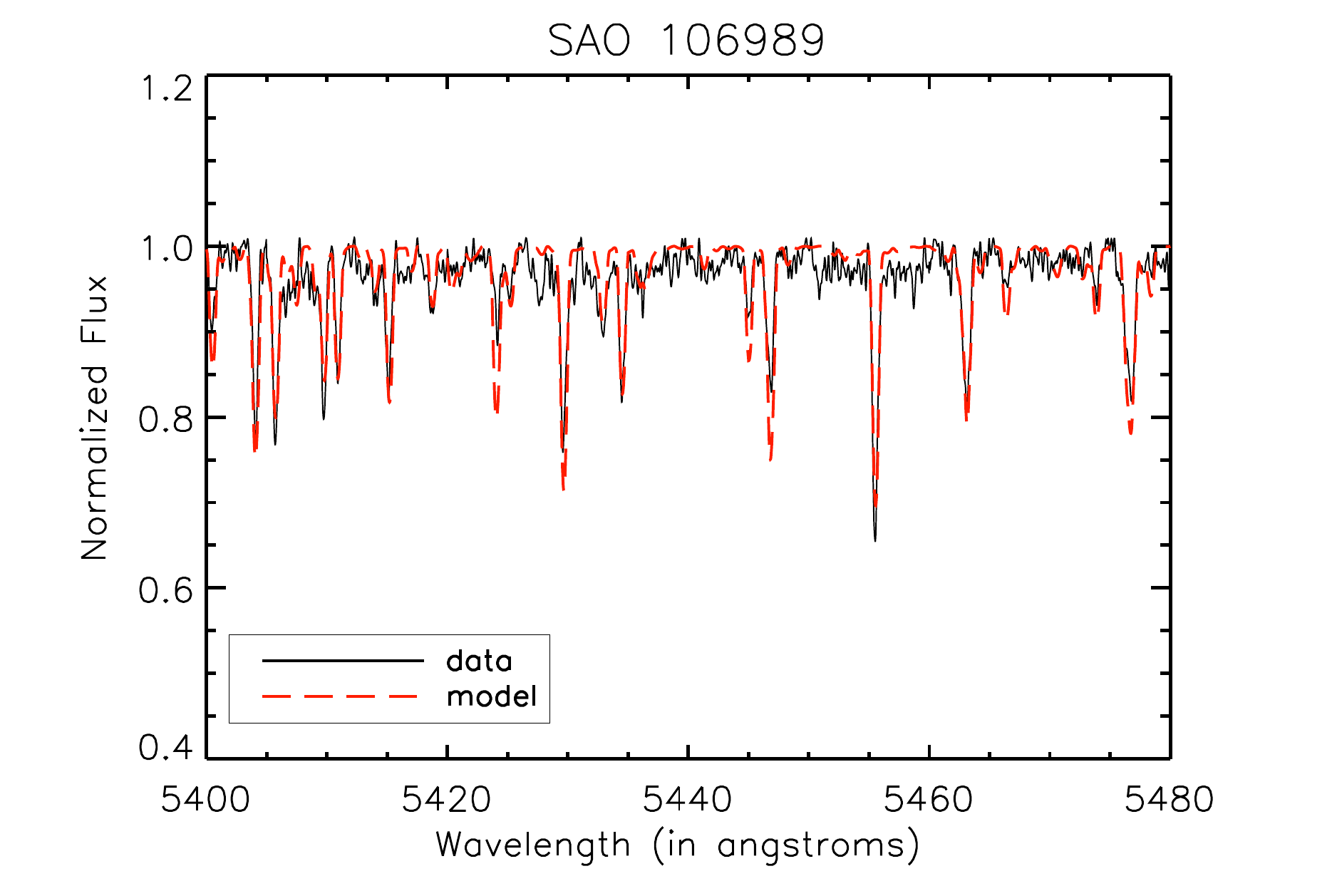}
\includegraphics[width=0.42\textwidth]{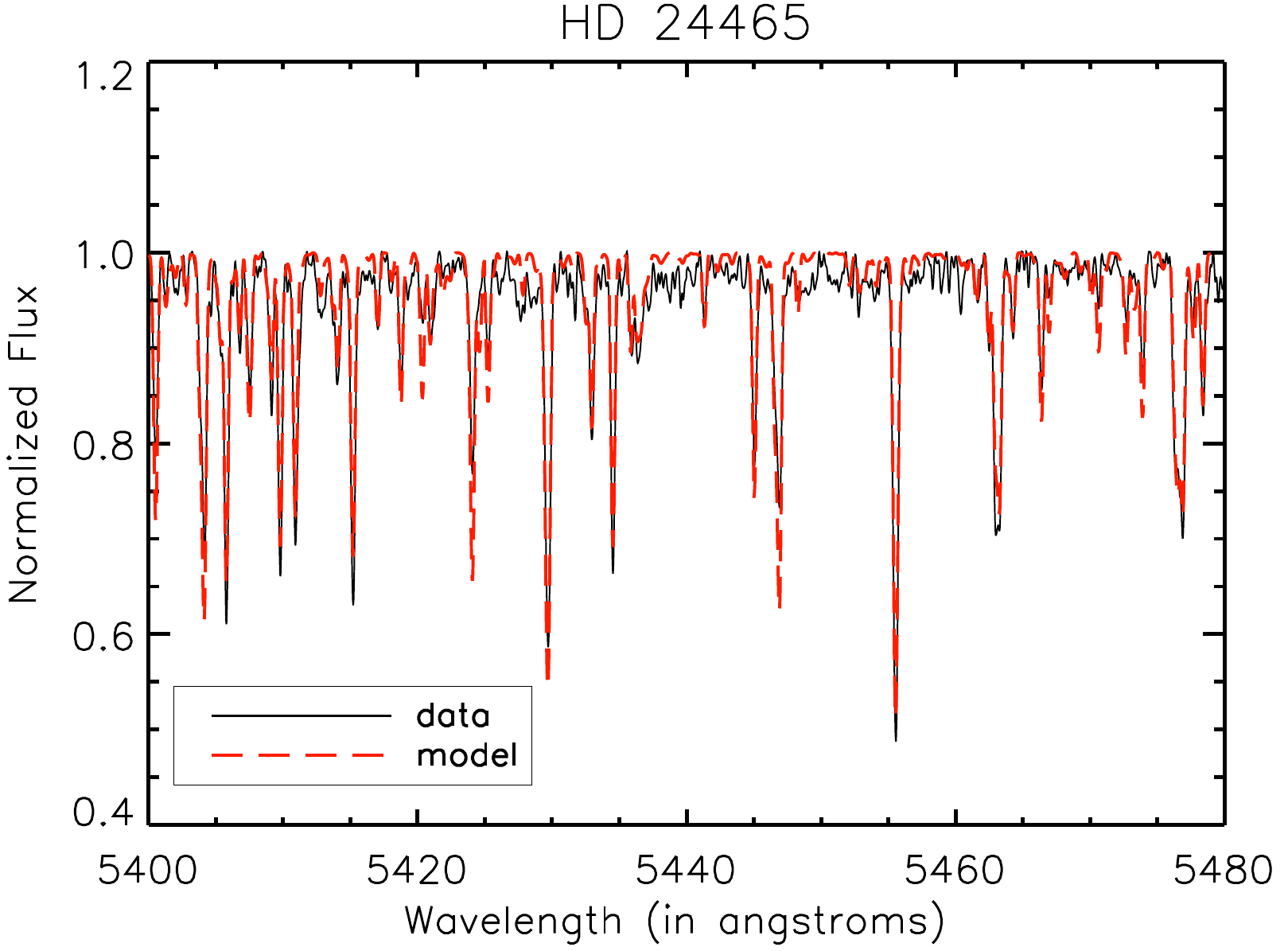} 
\includegraphics[width=0.48\textwidth]{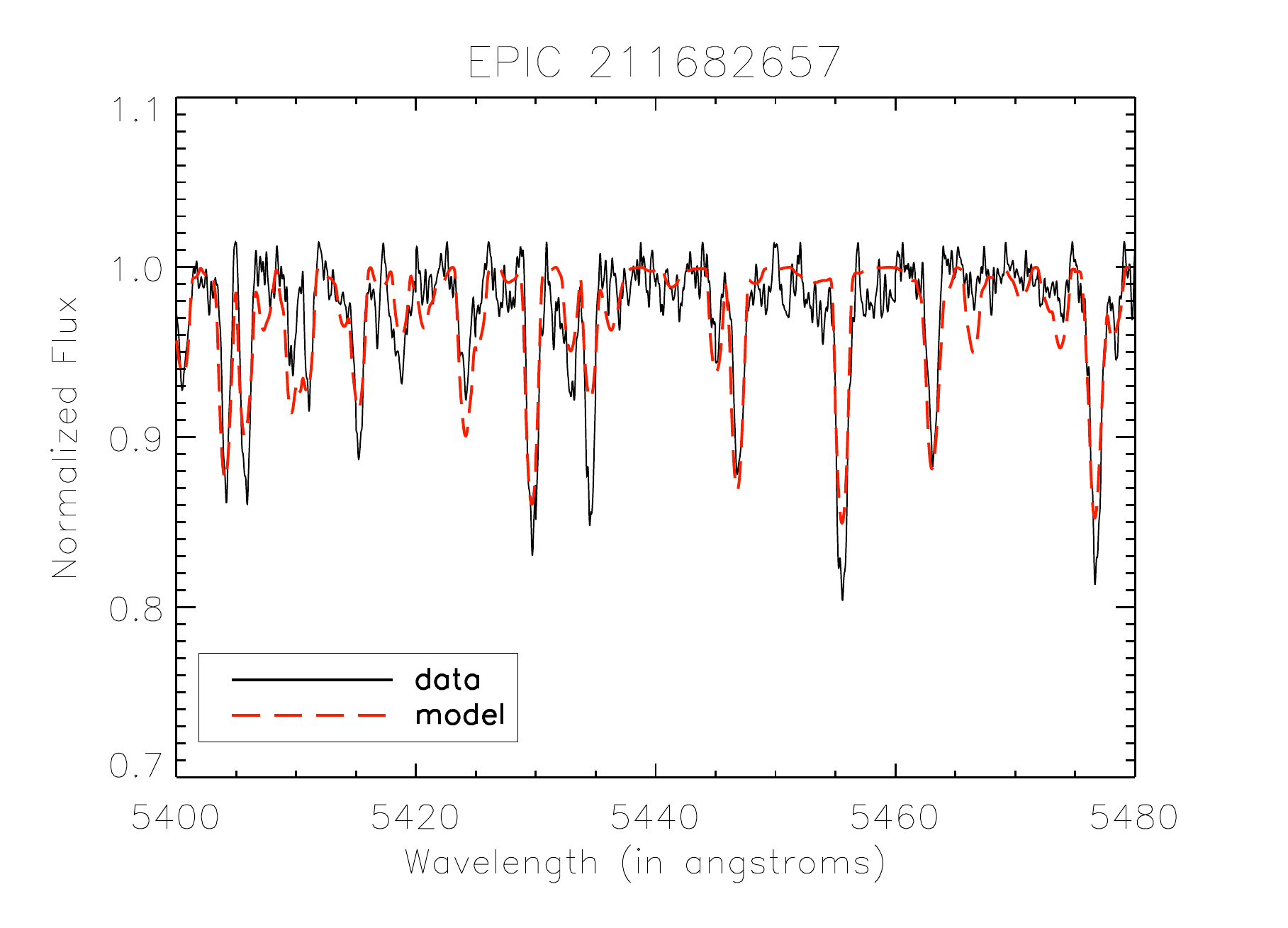} 
\includegraphics[width=0.49\textwidth]{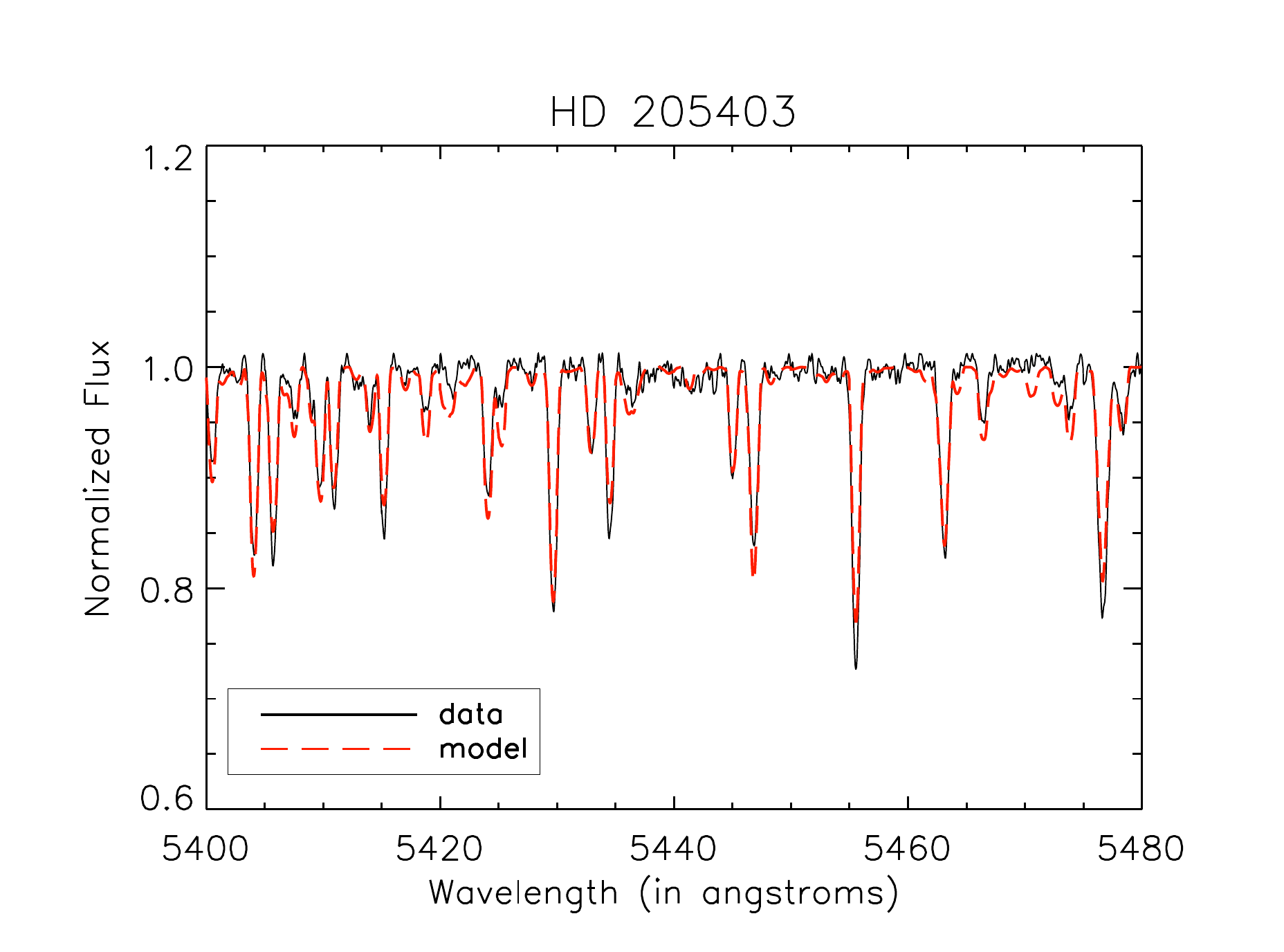} 
\caption{The solid line in each panel represents observed normalized spectra of SAO~106989 (top left), HD~24465 (top right), EPIC~211682657 (bottom left), and HD~205403 (bottom right) plotted across the wavelength region of $5400-5480~\AA$. Overplotted is the respective modelled spectra in dash line obtained from PARAS SPEC. The stellar parameters for each of the derived models are listed in Table~\ref{tab:paras_res}. For higher SNR, the spectra shown here is smoothed by 1.5 times leading to a resolving power of 44000. For details please refer text. \label{fig:spec_all} } 
\end{figure*}

\begin{figure}
\centering
\includegraphics[width=0.48\textwidth]{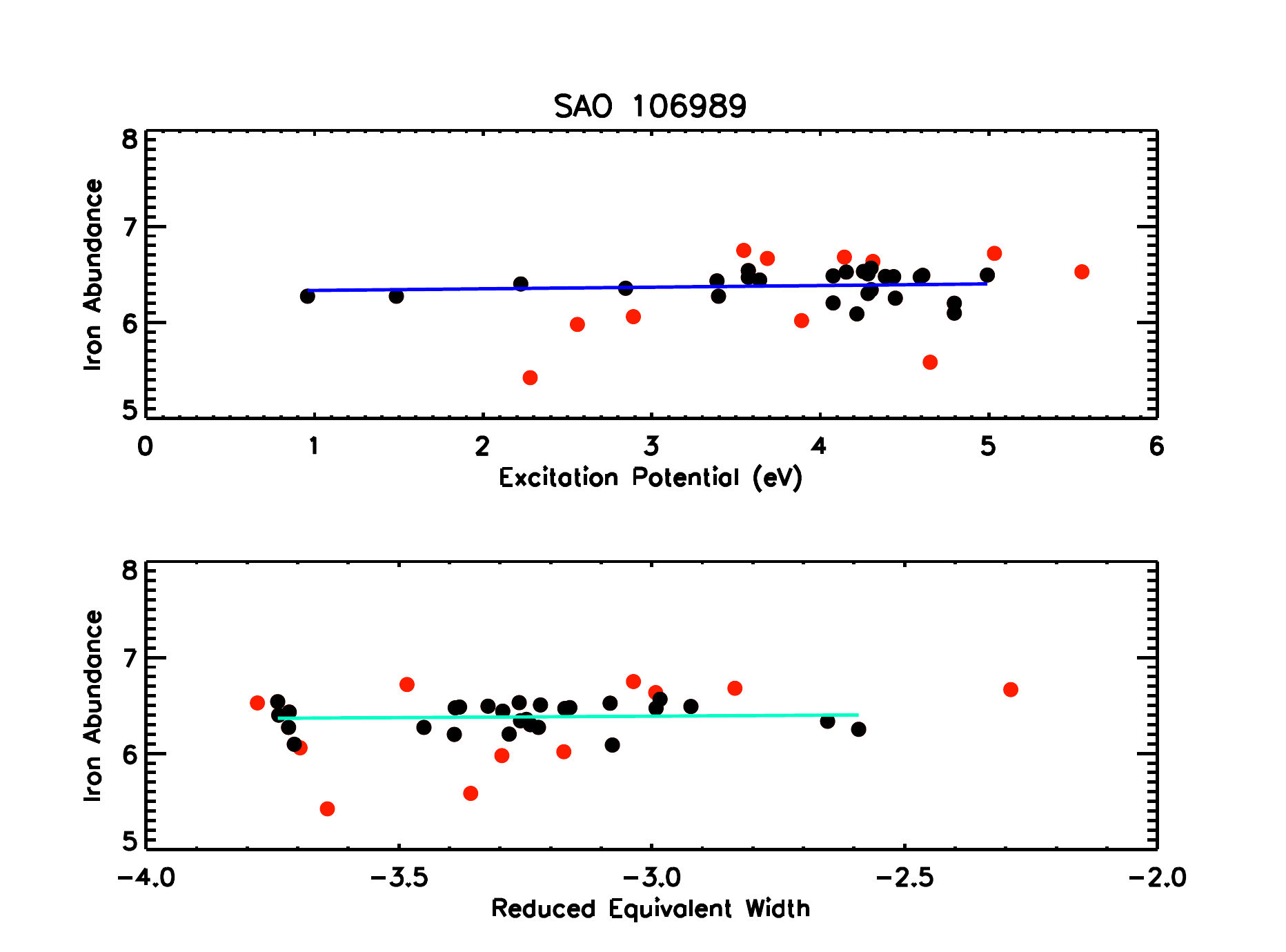}
\includegraphics[width=0.48\textwidth]{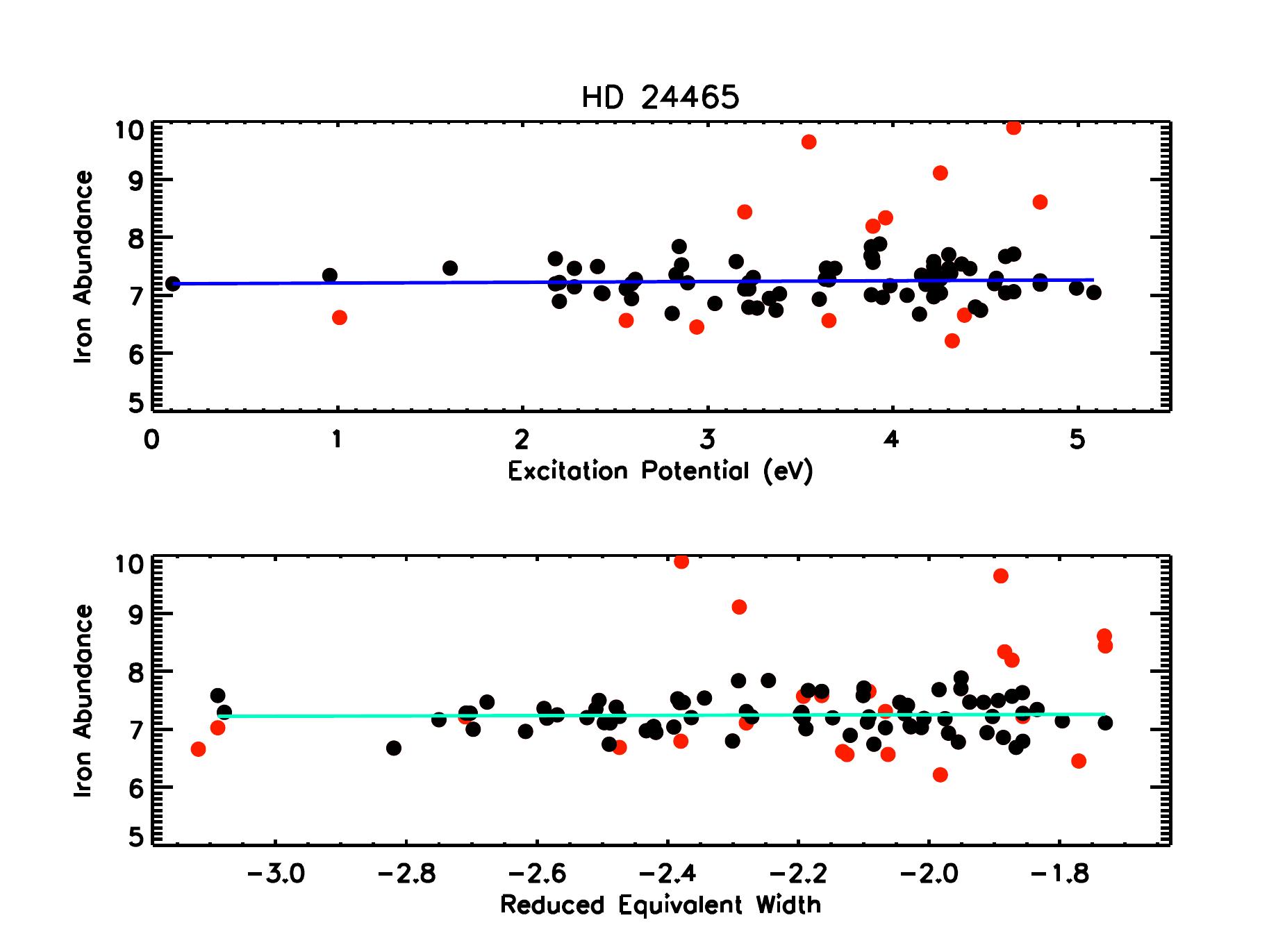} \\
\includegraphics[width=0.48\textwidth]{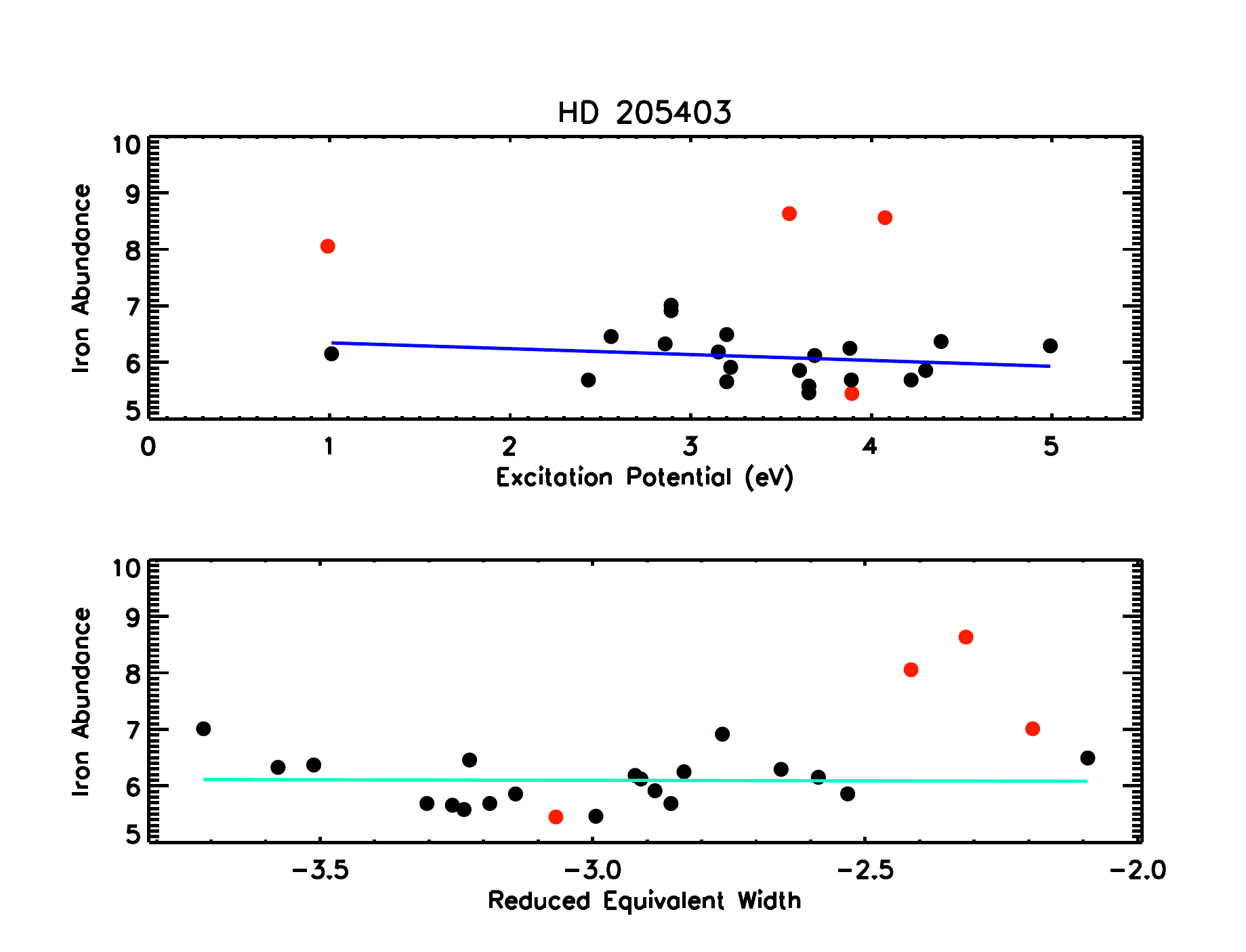}
\caption{The top sub-figure in each panel is iron abundance for SAO~106989 (top left), HD~24465 (top right), and HD~205403 (bottom) plotted against EP for each Fe I or Fe II line from the line list. The blue line is the least-square fit to each data point seen in the scatter plot indicating the minimum slope for the best-determined $T\rm{_{eff}}$. In the bottom sub-figure, iron abundance is plotted against reduced EW for each panel and the blue-green line indicates the minimum slope for the least-square fit obtained on the data for best determined $v_{\rm{micro}}$ for each of the stars. The red points are the discarded points having standard deviation beyond 1 $\sigma$ (not considered for the fit) The stellar parameters for each of the derived models are listed in Table~\ref{tab:paras_res}. \label{fig:ew_all}}
\end{figure}

%------------------------------------------------------------------------------------------------

\subsubsection{Synthetic spectral fitting method}

The synthetic spectral fitting method is a four step automated execution to determine $T_{\rm{eff}}$, $[Fe/H]$, ${\rm log}~g$ and $v sin i $. The RMS residuals ($\sum\nolimits {(O(i)-M(i))}^2$) are computed between the observed (O) and the modelled spectra (M) at each wavelength bin, $\lambda _i$, in the utilized wavelength region. The model producing the best-match with the observed spectra gives the best-fit values of $T_{\rm{eff}}$, $[Fe/H]$, ${\rm log}~g$ and $v sin i $. For the first step, all the parameters are kept free and maximum wavelength range ($5050-6500~\AA$) is used as there are many temperature and metallicity lines in this region. The best-fit values of $T_{\rm{eff}}$, $[Fe/H]$ and $v sin i $ are stored from this execution and used for the next steps. In the second step, $T_{\rm{eff}}$ and $[Fe/H]$ are kept frozen and used as an initial guess value whereas the value of ${\rm log}~g$ is kept free. The current step is executed only on the ${\rm log}~g$ sensitive Mg~I lines in the wavelength region of $5160-5190~\AA$. The initial two steps are executed on a coarse set of library. In order to get better precision on stellar parameters, a finer grid is required to achieve closest match between observed and synthetic spectra. Thus, during the course of execution of the synthetic spectral fitting routine, the synthetic models are interpolated in the desired range of $T_{\rm{eff}}$, $[Fe/H]$ and ${\rm log}~g$ to sharpen the precision of the derived parameters. The interpolation on the models is executed by the {\tt{IDL}} subroutine {\tt{kmod}}. The interpolated models then have a finer interval in $T_{\rm{eff}}$ (50 K), $[Fe/H]$ (0.1 dex) and ${\rm log}~g$ (0.1 dex). The third step is applied on interpolated models. The parameters obtained from the previous two steps are used as initial guess values and interpolation is done simultaneously at finer precision. The interpolation is done in the vicinity of the guess values on the three parameters obtained from the first step, i.e., $T_{\rm{eff}}$ in a range of $\pm$~250 K, $[Fe/H]$ in a range of $\pm$ 0.3 and ${\rm log}~g$ in a range of $\pm$ 0.3. The best-determined values of $T_{\rm{eff}}$ and $[Fe/H]$ derived from this second step are considered as initial approximations on stellar parameters for the last step. The last step is executed for determining the ${\rm log}~g$ from the wavelength region of $5160-5190~\AA$ on the interpolated models. The $T_{\rm{eff}}$ and $[Fe/H]$ derived from the third step are used in this last step. This step is similar in execution to the second step, the only difference being is that it is being applied on an interpolated finer grid. The best-match model determined at this step gives us the value for ${\rm log}~g$ along with previously determined values of $T_{\rm{eff}}$ and $[Fe/H]$ from the third step.

For the synthetic spectral fitting method to work, one needs a high SNR ($\geq80$) observed spectra. Since all the program stars studied as a part of this work are of F spectral type, similar procedures have been adopted for estimating the stellar parameters. We combined all 17 observed epochs (See Table~\ref{tab:rv_sao}) for SAO~106989. The SNR/pixel in the case of the star SAO 106989 for the combined 17 epochs was $\sim 80-85$ in the wavelength region between $6000-6500$~\AA~and $55-75$ in the wavelength region of $5000-6000$~\AA. There are prominent metallic lines in the blue region of the spectra for a F type star. However, this wavelength region of the spectra have less SNR and thereby in order to effectively use this wavelength region, we smoothed the co-added spectra by a factor of 1.5. This enhanced the SNR to $80-100$ in the blue region at a resolving power of $\sim$~44000. Thereby, the wavelength region $5200-5700~\AA$~of the spectra was used for synthetic spectral fitting method.  The top left panel of Fig.~\ref{fig:spec_all} shows a sample of the observed spectrum (solid line) overlaid by the best-fit model in dotted line across the wavelength region $5400-5480~\AA$. The best-fit stellar parameters for the spectra derived from this method are $T_{\rm{eff}}$~=$6000\pm100$~K, $[Fe/H]=-0.2\pm0.1$ and ${\rm log}~g=4.2\pm0.2$. Similarly, the 14 epochs observed for HD~24465 (See Table~\ref{tab:rv_sao}) were combined and smoothed by a factor of 1.5 for an increased SNR. The combined SNR/pixel for HD~24465 was found to be $\sim 80-90$ in the wavelength region of 6000-6500~\AA~and between $60-80$ in the wavelength region of 5000-6000~\AA. We used the same wavelength region of $5200-5700~\AA$~for the synthetic spectral fitting. The top right panel of Fig.~\ref{fig:spec_all} shows a sample of the observed spectrum for HD~24465 (solid line) overlaid by the best-fit model in dotted line. The best-fit derived parameters by \textit{PARAS SPEC} for the model are $T_{\rm{eff}}$~=~$6250\pm100$~K, $[Fe/H]=0.3\pm0.15$, and ${\rm log}~g=4.0\pm0.15$. For EPIC~211682657, we found the SNR/pixel to be $\sim 70-80$ and $65-70$ in the wavelength region between $5000-6000$~and $6000-6500$~\AA~respectively after combining the data for all the observed 18 epochs (See Table~\ref{tab:rv_sao}) and smoothening to a resolving power of $\sim$~44000. Applying the same routine, we derive the stellar parameters for this star as $T_{\rm{eff}}$~=~$6650\pm125$~K, $[Fe/H]=-0.1\pm0.15$ and ${\rm log}~g=3.8\pm0.15$. The bottom left panel of Fig.~\ref{fig:spec_all} shows observed spectrum for EPIC~211682657 (solid line) overlaid by the best-fit model (dotted line). Finally, we combined all the available 21 epochs for HD~205403 (See Table~\ref{tab:rv_sao}) and smoothed the spectra, which resulted in SNR/pixel $\sim 80-90$ in the wavelength region between $5000-6000$~\AA~and $75-80$ in the wavelength region of $6000-6500$~\AA. The best-fit derived stellar parameters from \textit{PARAS SPEC} are $T_{\rm{eff}}$~=~6600~K, $[Fe/H]=-0.1$ and ${\rm log}~g=3.5$. The bottom right panel of Fig.~\ref{fig:spec_all} shows a sample of the observed spectrum (solid line) overlaid by the best-fit model in dotted line. The SNR/pixel of the combined spectra for all the stars studied here are in general below 100, which causes continuum matching errors as discussed in \cite{Chaturvedi2016}. Thus, the errors on each stellar parameter are found to be relatively larger, of the order of $\pm$75-125~K for $T\rm{_{eff}}$ and $\pm$0.1-0.15~dex for $[Fe/H]$ and ${\rm log}~g$. 

\subsubsection{Equivalent width method}

The Equivalent method (EW hereafter) was used in order to check and verify results obtained from the synthetic spectral line fitting method \citep{Blanco-Cuaresma2014}. It works on the principle in which one seeks the neutral and ionized iron lines to satisfy the two equilibria, namely, excitation equilibrium and ionization balance. A set of neutral and singly ionized lines is acquired from the iron line list by \cite{Sousa2014}. Identification of unblended lines for determination of equivalent widths (EW) is a pre-requisite for this method. For this method, similar to the previous method, we utilized the combined higher SNR/pixel observed spectra. The {\tt{SPECTRUM}} code facilitates estimation of abundance of elements from their spectral lines by using a set of EW of the fitted lines as an input to the {\tt{ABUNDANCE}} subroutine. The subroutine also uses various stellar models which are formed as a combination of different $T\rm{_{eff}}$, $[Fe/H]$, ${\rm log}~g$ and $v_{\rm{micro}}$. The main purpose of calculating EW and thereby abundances is the fact that the abundances of a given species follow a set of three golden rules \citep{Neves2009,Blanco-Cuaresma2014}. This fact can be exploited to choose a best-fit model of synthetic spectra in which all the rules are simultaneously satisfied. These three rules are:
\begin{itemize} [noitemsep,topsep=0pt]
\item Abundances as a function of excitation potential (EP) should have no trends. 
\item Abundances as a function of reduced EW (EW/$\lambda)$ should exhibit no trends.
\item Abundances of neutral iron (Fe I) and ionized iron (Fe II) should be balanced. 
\end{itemize}
For each iteration, abundances are calculated as a function of a set of stellar parameters ($T\rm{_{eff}}$,  ${\rm log}~g$ and $v_{\rm{micro}}$). The derived abundances are plotted as a function of EP and reduced EW. Slopes are fit to these plots by fitting a linear polynomial. A difference of Fe I and Fe II abundances is also calculated for each set of stellar parameters. 
Both the parameters, $T\rm{_{eff}}$ and $v_{\rm{micro}}$ are determined simultaneously by minimized slopes as mentioned previously. A slight positive or negative slope indicates under-estimation or over-estimation of $T_{\rm{eff}}$ and $v_{\rm{micro}}$ for the star respectively. Similarly, if the Fe I and Fe II difference is positive or negative, it indicates that ${\rm log}~g$ is under-estimated or over-estimated respectively. The entire process is executed in two steps: first step on the coarse grid of models in $T\rm{_{eff}}$, ${\rm log}~g$ and $v_{\rm{micro}}$ and second step on the interpolated finer grid, similar to the previous method of synthetic spectral fitting. Thus, the model having a set of parameters where the slopes of iron abundances against EP and reduced EW and the differences between neutral and ionized iron abundances are simultaneously minimized gives us the best-determined $T\rm{_{eff}}$, ${\rm log}~g$ and $v_{\rm{micro}}$.

SAO~106989 has a magnitude of 9.3 and is towards the fainter limit of observations for PARAS. The star has a relatively large rotational velocity (20~km~s$^{-1}$) leading to blending of closely situated lines. Thus, there are fewer number of unblended Fe~I and Fe~II lines identified for abundance determination by EW method. 
%The list of unblended spectral lines considered for analysis by the EW method are given in Table~\ref{tab:ew_lines_sao106989}. 
In the top left panel of Fig.~\ref{fig:ew_all}, a least-square fit having a minimum slope for iron abundances vs. excitation potential (EP) obtained for best-fit $T_{\rm{eff}}$ for SAO~106989 is shown in the upper sub-figure of the panel. In the bottom sub-figure, a plot of iron abundance vs reduced EW is shown with a least-square fit line having a minimum slope for the best-fit $v_{\rm{micro}}$. 
%The spectral properties determined by both the methods for SAO~106989 are listed in Table~\ref{tab:paras_res}.
HD~24465 has a magnitude of 8.9 and a rotational velocity (11~km~s$^{-1}$). The number of Fe~I and Fe~II lines identified for abundance determination through EW were sufficient as compared to SAO~106989. 
%The list of unblended spectral lines considered for analysis by the EW method are given in Table~\ref{tab:ew_lines_hd24465}. 
In the top right panel of Fig.~\ref{fig:ew_all}, a least-square fit line having a minimum slope for iron abundances vs. EP obtained for best-fit $T_{\rm{eff}}$ for HD~24465 is shown in the upper sub-figure. In the bottom sub-figure, a plot of iron abundance vs reduced EW is shown with a least-square fit line having a minimum slope for the best-fit $v_{\rm{micro}}$. %The spectral properties determined by both the methods for HD~24465 are listed in Table~\ref{tab:paras_res}.
 EPIC~211682657 has a very large rotational velocity of 40~km~s$^{-1}$. Thus, there were no unblended lines available for measurement of EW to determine stellar parameters and hence the EW method could not be used for this star. The rotational velocity for HD~205403, is 16~km~s$^{-1}$. 
%The list of lines considered for the EW method are given in Table~\ref{tab:ew_lines_hd205403}. 
In the bottom panel of Fig.~\ref{fig:ew_all}, a least-square fit having a minimum slope for iron abundances vs. EP obtained for best-fit $T_{\rm{eff}}$ for HD~205403 is shown in the upper sub-figure of the panel. In the bottom sub-figure of the panel, a plot of iron abundance vs reduced EW is shown with a least-square fit line having a minimum slope for the best-fit $v_{\rm{micro}}$. The spectral properties determined by both the methods for all the program stars are listed in Table~\ref{tab:paras_res}. We can see from Table~\ref{tab:paras_res} that spectral parameters derived from both these methods agree within the uncertainities given in the Table. We have used the average value of the parameters derived by these two methods for further analysis.

%-------------------------------------------------------------------------
% TABLE 2
%-------------------------------------------------------------------------
\startlongtable
\begin{deluxetable}{lcc}
\small	
\tablecaption{Spectral properties of all the primary stars derived by \textit{PARAS SPEC}. Note that the EW method could not be applied for EPIC~211682657. For details, please refer text. \label{tab:paras_res}}
\tablehead{
\colhead{Parameters} & \colhead{Synthetic spectral fitting} & \colhead{EW method} \\
}
\startdata
\textbf{SAO~106989} && \\
$T\rm{_{eff}}$ (K) & $6000\pm100$ & $5925\pm100$\\
$[Fe/H]$ & $-0.2\pm0.1$ & $-0.2$ (fixed)\\
${\rm log}~g$ & $4.2\pm0.2$ & $4.25\pm0.1$\\
v$_{\rm{micro}}$ (km~s$^{-1}$) & -- & $0.5\pm0.1$\\
$v \sin i$ (km~s$^{-1}$) & $20\pm2$ & -- \\
\textbf{HD~24465} && \\
$T\rm{_{eff}}$ & $6250\pm100$ & $6150\pm75$\\
$[Fe/H]$ & $0.3\pm0.15$ & $0.3$ (fixed)\\
${\rm log}~g$ & $4.0\pm0.15$ & $4.06\pm0.1$\\
v$_{\rm{micro}}$ (km~s$^{-1}$) & -- & $0.5\pm0.1$\\
$v \sin i$ (km~s$^{-1}$) & $11\pm1$ & -- \\
\textbf{EPIC~211682657} && \\
$T\rm{_{eff}}$ & $6650\pm125$ \\
$[Fe/H]$ & $-0.1\pm0.15$ \\
${\rm log}~g$ & $3.8\pm0.15$ \\
$v \sin i$ (km~s$^{-1}$) & $40\pm1$ \\
\textbf{HD~205403} && \\
$T\rm{_{eff}}$ & $6600\pm100$ & $6450\pm75$\\
$[Fe/H]$ & $-0.1\pm0.15$ & $-0.1$ (fixed)\\
${\rm log}~g$ & $3.5\pm0.15$ & $3.7\pm0.1$\\
v$_{\rm{micro}}$ (km~s$^{-1}$) & -- & $1.4\pm0.1$\\
$v \sin i$ (km~s$^{-1}$) & $25\pm1$ & -- \\
\enddata
\end{deluxetable}

%-------------------------------------------------------------------------
             
\subsection{Photometry of all the primary stars of the EB systems}

In this section, we describe the retrieval and analysis of the archival data for each of the sources.

%------------------------------------------------------------------------------------------------
%  FIG 4
%------------------------------------------------------------------------------------------------
\begin{figure}
	\includegraphics[width=0.5\textwidth]{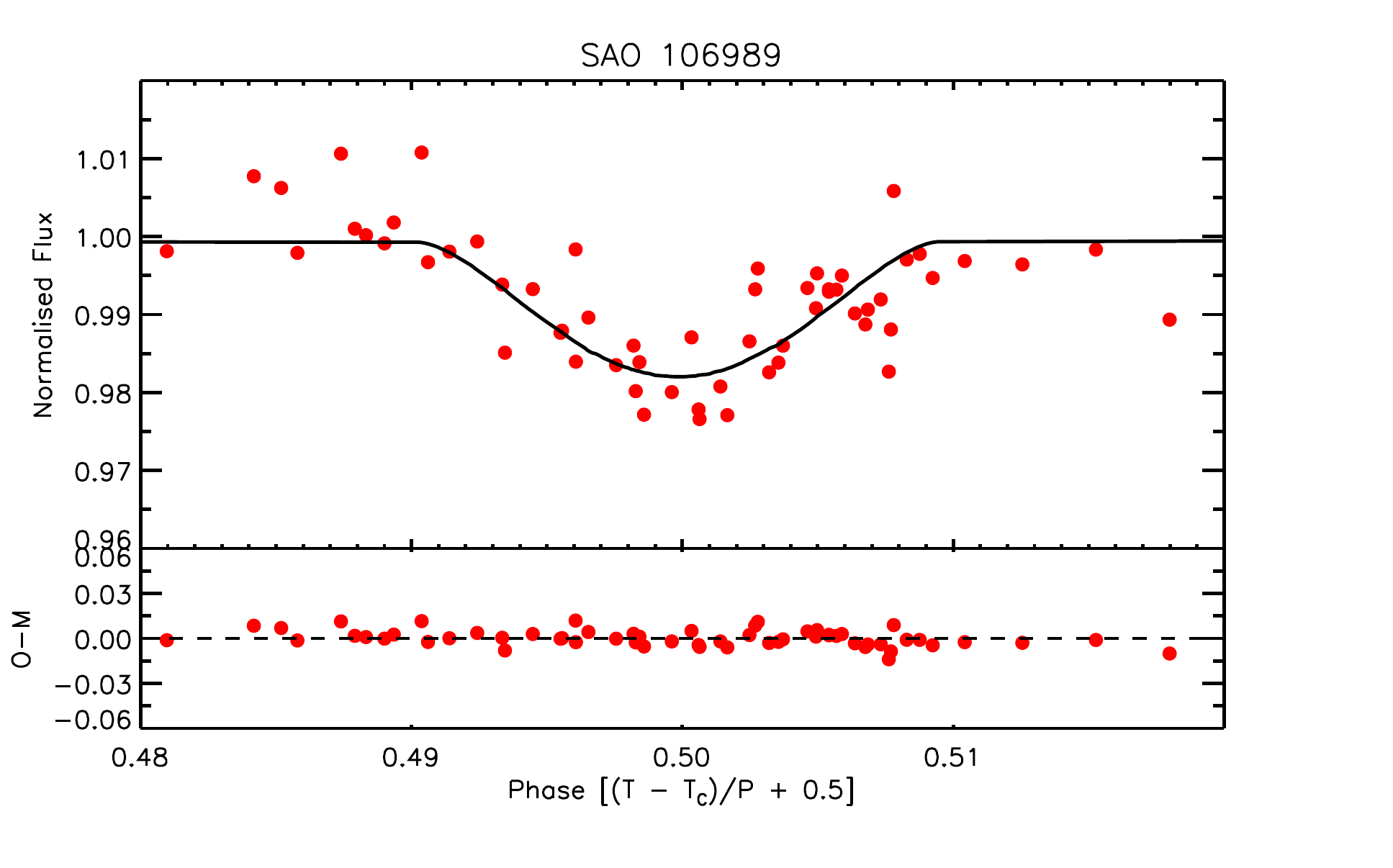}
	\includegraphics[width=0.5\textwidth]{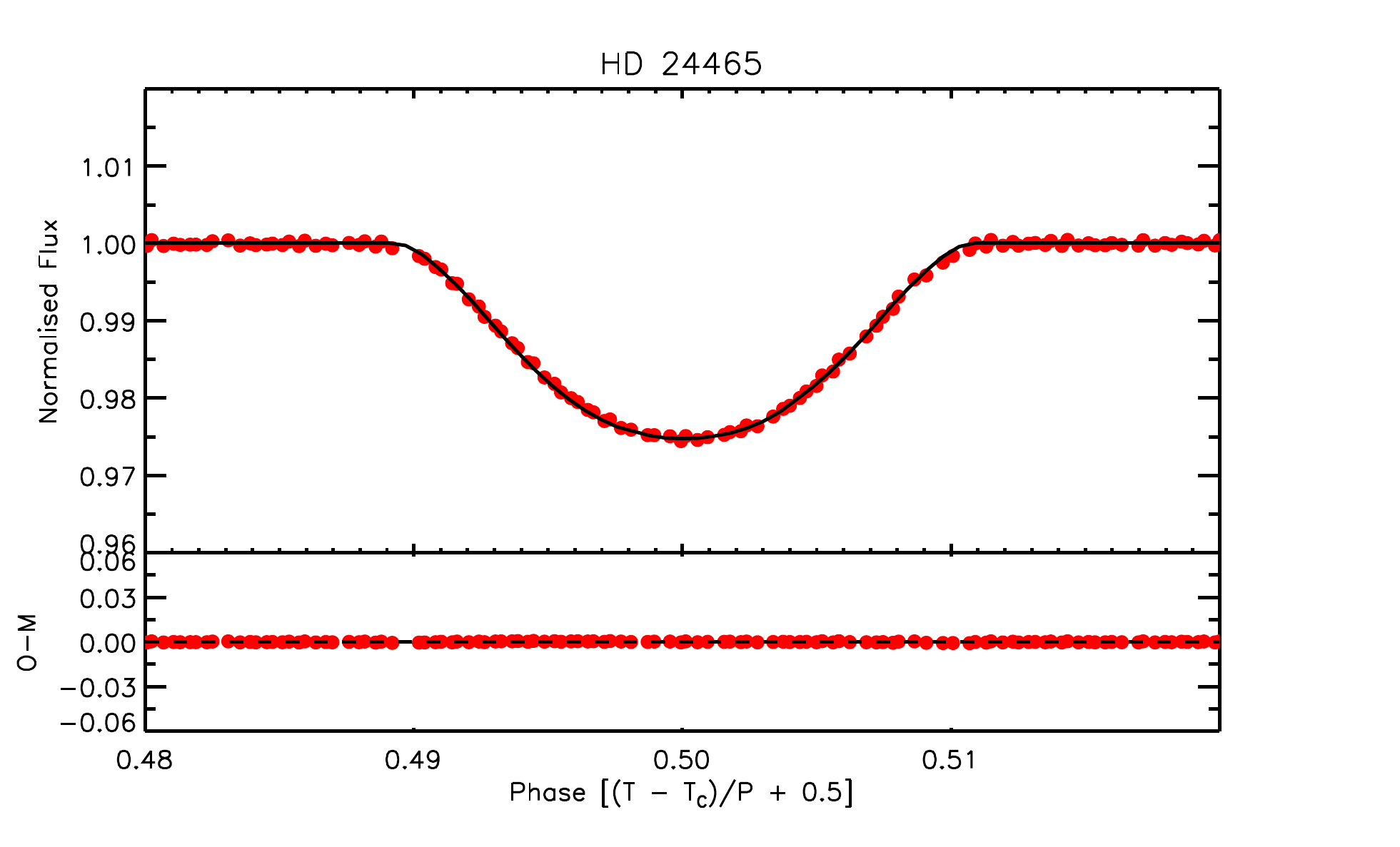}\\
	\includegraphics[width=0.5\textwidth]{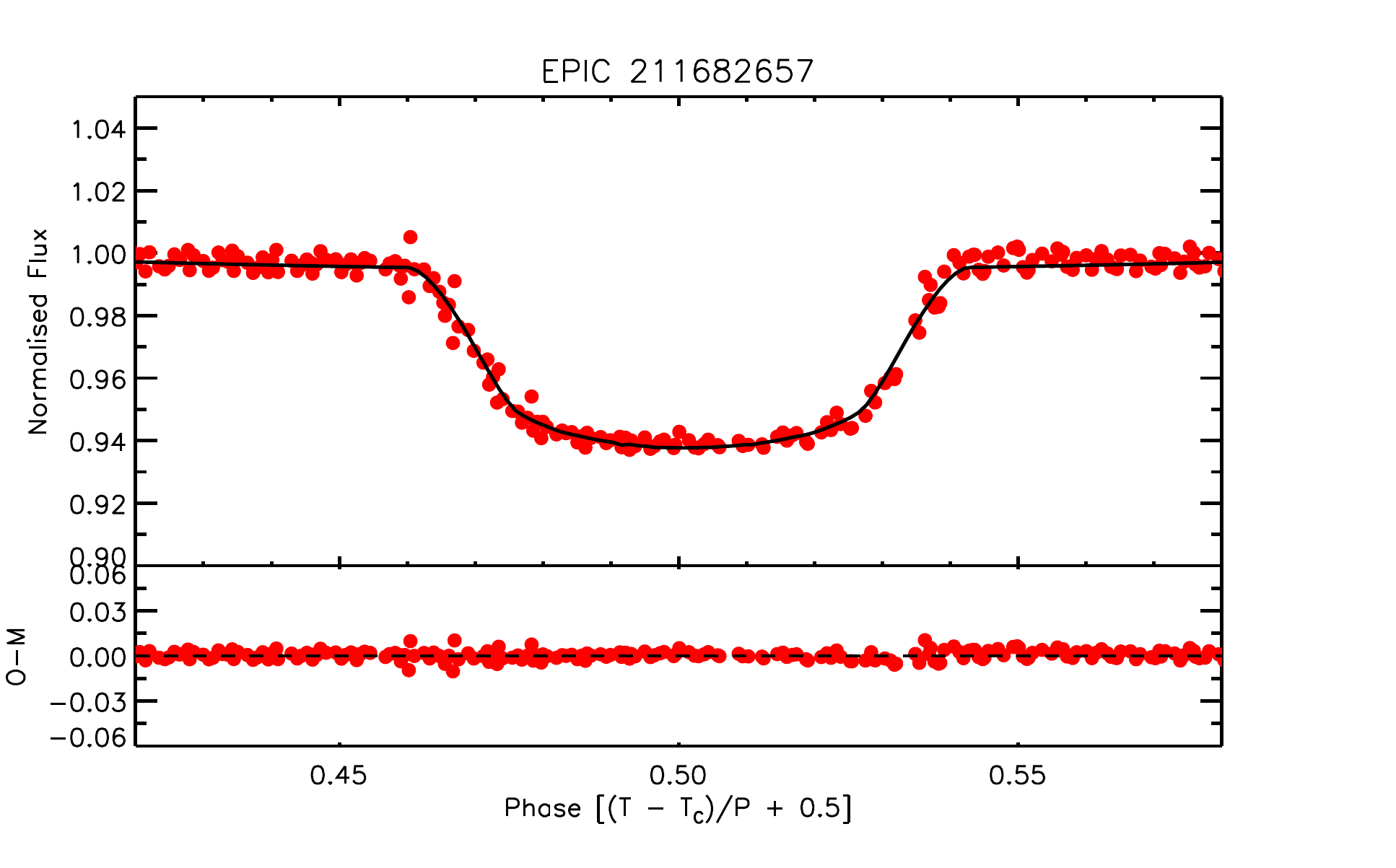} 
	\includegraphics[width=0.5\textwidth]{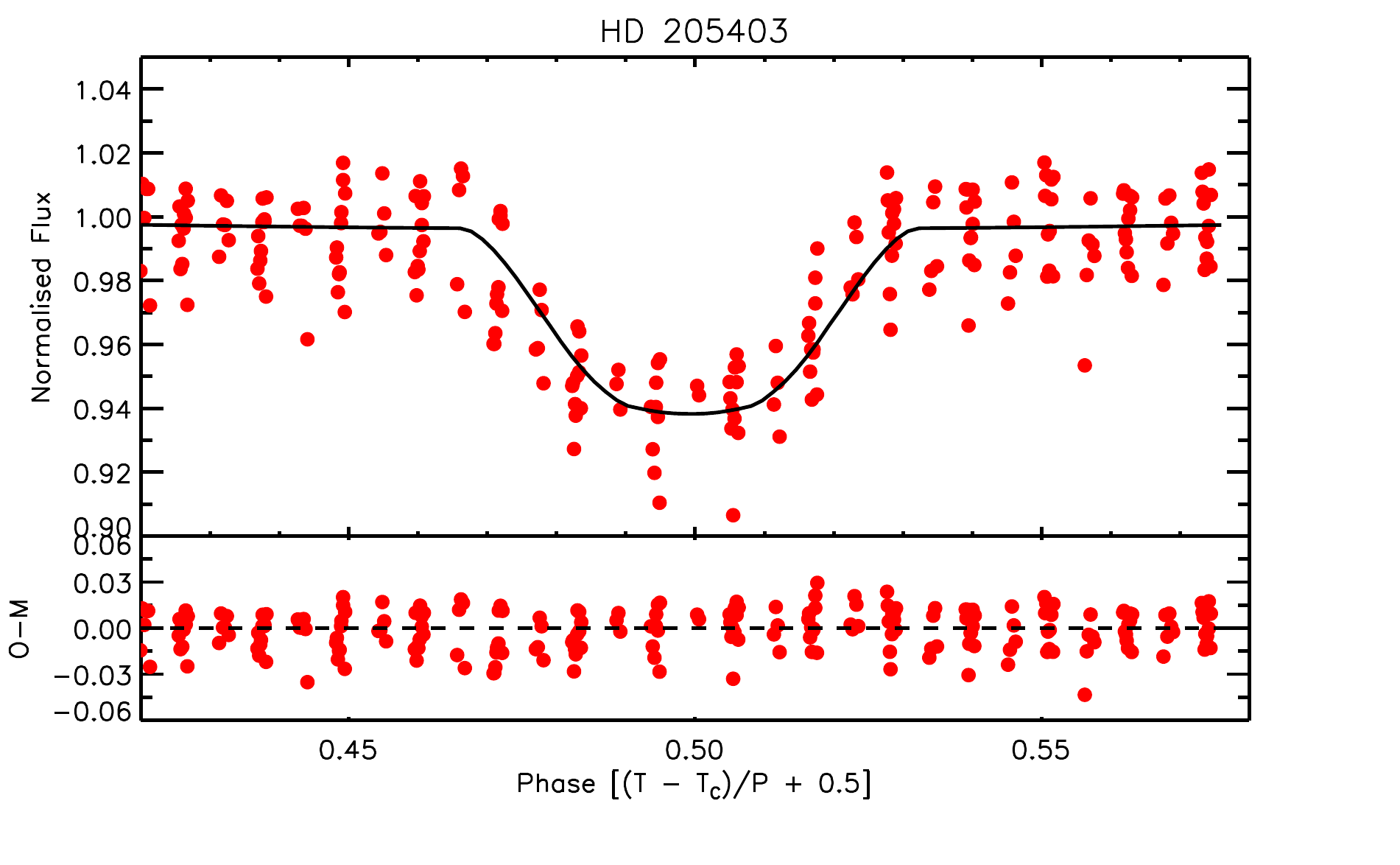}
\caption{(Top panel of each subfigure) Transit curve (filled circles) for star SAO~106989 obtained from KELT data (top left), for star HD~24465 obtained from K2 data (top right), for star EPIC~211682657 obtained from K2 data (bottom left) and HD~205403 obtaned from STEREO data are plotted based on the parameters derived from PHOEBE with a solid line. (Refer \S~\ref{subsec:orbital_sao} for details on PHOEBE)
(Bottom panel of each subfigure ) Observed-fit residuals are plotted. For better visual representation, the x-axis in phase is shifted by 0.5 so that the central primary transit crossing point (Tc) occurs at phase 0.5 instead of 0. \label{fig:trans_sao}}
\end{figure}

%------------------------------------------------------------------------------------------------

 We retrieved the reduced data for SAO~106989 from the photometry archives. All the photometry data available for $\sim~85$ nights between 8 June 2007--21 November 2008 was combined to reflect the transit signature as there were only partial eclipses recorded. Despite the KELT data being noisy for this source, we could fit the light curve at the same period as that for RV data as shown in the top left panel of Fig.~\ref{fig:trans_sao}.

We retrieve the {\tt{K2SFF}} (K2 Self Flat Fielding Correction) light curve data from MAST Portal \footnote{https://mast.stsci.edu/portal/Mashup/Clients/Mast/Portal.html} for the sources HD~24465 (EPIC~210484192) and EPIC~211682657 from \cite{Barros2016}. {\tt{K2SFF}} light curve data is publically available. K2 data is particularly noisy as compared to its predecessor Kepler. The technique of aperture photometry and imaging centroid position is applied to account for spacecraft's motion. This technique incorporates for the non-uniform pixel response function of the K2 detectors by correlating the measured flux with spacecraft's pointing angle and correcting for such dependence \citep{Vanderburg2014}. Datapoints with poor photometric performance are removed and variabilities of the order of 6 h caused by spacecraft jitter are also removed. B spline function is fitted iteratively to the datapoints in order to eliminate low frequency variability. The acquired data for both the stars are long cadence data with an average time cadence of 30 min. We could detect 23 complete eclipses for EPIC~211682657 and 10 eclipses for HD~24465 for $\sim~75$ nights of observations for each of the fields. The photometry data and the period determined by RV data agree well with each other. The transit data fitted at the orbital period of 7.197 and 3.142~d for HD~24465 and EPIC~211682657 are shown respectively in the top right and bottom left panels of Fig.~\ref{fig:trans_sao}.

For the EB HD~205403, STEREO data from HI-1A instrument were extracted from the UK Solar System Data Centre (UKSSDC) website~\footnote{www.ukssdc.rl.ac.uk}. We extracted the available bias-subtracted and flat-fielded data between Dec 2008-Nov 2010, which constituted 8 complete transit events for $\sim~35$ nights of observations for the two cycles. The spacecraft coordinates were converted to sky coordinates to identify the star. We used the standard IRAF~\footnote{IRAF is distributed by the National Optical Astronomy Observatory, which is operated by the Association of Universities for Research in Astronomy,
Inc., under cooperative agreement with the National Science Foundation.} DAOPHOT package for processing the photometry data and continuum normalized for light curve fitting. Aperture photometry was applied for different aperture sizes of 2.5, 3.0 and 3.5 pixels around the star. A larger aperture included too much background contribution and a smaller aperture had the star light spill over the aperture in some of the frames. Thus, an aperture of 3 pixels was found appropriate. Sky background was calculated between pixel radii 7.0 and 10 as there was no contamination from neighbouring sources. Photon-electron conversion gain for the camera was kept at 15 units \citep{Sangaralingam2011}. The rms scatter on the light curve for the source star outside the transit time duration is 7 mmag. The photometry data fitted and phase-folded at a period of 2.44~d and our matched RV derived period, is shown in the bottom right panel of Fig~\ref{fig:trans_sao}. We also searched for the secondary eclipse but did not find any significant evidence. The secondary eclipse depth for all the sources were either undetectable or small. The data for SAO~106989 is not modeled by PHOEBE for secondary eclipse as the data was significantly noisy. The secondary eclipse depths for HD~24465 and EPIC~211682657 as modeled from KEPLER data are 0.000018 and 0.009 (in normalized flux units) respectively. For the source HD~205403, the secondary eclipse depth is 0.0088 but the standard deviation for out of transit points is 0.0108 and is thus undetectable.

\subsection{Orbital parameters for SAO~106989, HD~24465, EPIC~211682657 $\&$ HD~205403} \label{subsec:orbital_sao}

We utilized PHOEBE (PHysics Of Eclipsing BinariEs) routine \citep{Prsa2005} for the modelling of the light curves and radial velocity data for the four EBs, SAO~106989, HD~24465, EPIC~211682657 and HD~205403. The routine is based on the WD’s differential corrections method of Wilson $\&$ Devinney \citep{Wilson1971} using the Nelder $\&$ Mead’s downhill simplex for minimization based on function evaluations. The routine reads the photometry and RV data, a set of initial parameters depending on the physical and geometrical properties of the system, and the minimization algorithm needed for the process. PHOEBE has a back-end scripter that can facilitate the implementation of heuristic scans
of the solutions to probe the parameter degeneracies and avoid local minima \citep{Gomez2010}. Heuristic scanning offers improvisation to minimization algorithms by selecting starting points in parameter hyperspace and minimizing from each point. The obtained parameters are weighted appropriately by sorting solutions based on cost function. Simulated annealing is another method used to avoid local minima \citep{Prsa2005}.

We have extensively referred to \cite{Gomez2010} thesis for developing a methodology to extract optimum system parameters for the EBs in consideration. A model-- detached, contact or semi-contact binary for the system has to be chosen in the interface menu based on understanding of general physics and geometry of the system. Stellar surfaces are considered as equipotential surfaces best described by Roche model. The surface potentials determine the shape and size of the components \citep{Kallrath2009}. The algorithm initially uses only RV data to fix RV dependent orbital parameters. The mid-transit time `T$_0$', the orbital period `P' and angle of inclination `i' is kept fixed for this iteration. The RV data is fitted independently to constrain the mass ratio `q', semi-RV amplitude `K', semi-major axis as a function of angle of inclination `a~sini', eccentricity `e', angle of periastron `$\omega$', phase shift `$\phi$', and line of sight velocity `$v{_\gamma}$' by using DC minimization. A single iteration gives some solution parameters that are returned for the user to inspect. Each time these parameters are resubmitted to improve the quality of the RV fit thereby minimizing the cost function ($\chi^2$) of the solution. The cost function converges and free parameters do not change within the error limits. This is the time one stops iterating the system any further. The values obtained from the RV iteration are noted and used for further analysis. The second part involves fitting of lightcurve (LC) data keeping the RV obtained parameters fixed. PHOEBE supports linear, logarithmic, and square root limb-darkening (LD) laws. We utilize the logarithmic ones for the case of optical wavelength regime, as it suits the best. The LD coefficients are modified dynamically by using \cite{Hamme1993} tables and are linearly interpolated to obtain appropriate values. The primary temperature of the system `T$_{\rm{eff}_1}$' is kept fixed derived from spectral analysis. The parameters, `i', primary surface potential $\Omega_1$, secondary surface potential $\Omega_2$, and secondary temperature `T$_{\rm{eff}_2}$' are kept free for fitting. 

We fixed the albedo and gravity brightening coefficients at 0.5 and 0.32 respectively for both components as both the primary and secondary stars here are having convective envelopes (T$_{\rm{eff}}$ $<$ 7200 K) \citep{Zasche2016}. We also assumed both components of the system to be synchronous. Similar to the RV iteration, LC iteration is executed till the cost function is minimized within error bars of the free parameters. Post each iteration, the value of `asini' is updated. The parameters `e', `$\omega$', and `$\phi$' depend on both RV and LC data. Finally, RV and the photometry data is fitted simulataneously to obtain a single consistent solution. The error bars on each derived quantity is obtained by method of error propagation as discussed in \cite{Gomez2010}. The values along with their respective error bars obtained for the orbital parameters are summarized in Table~\ref{tab:result_sao}. The error bars are estimated using linear propagation of errors. We have flagged the values in the Table~\ref{tab:result_sao} that are derived using error propagation. These parameters for which error propagation is used to estimate uncertainties, are derived parameters and not fitted parameters. The formal errors derived by us on mass and radius range from $1-3\%$ and $0.5-3\%$ respectively. These numbers when compared to the intrinsic scatter on mass and radius as seen from \cite{Torres2010} relation are $6.4\%$ and $3.2\%$ respectively. Thus, we see that formal uncertainties are smaller than the scatter in the Torres relation. This implies that the intrinsic scatter in the Torres relation dominates the uncertainty on mass and radius derived by measurements. It will be imperative in near future to work towards reducing this scatter in the empirical Torres relation for more reliable estimates of masses and radii of EB components. It is important to note that the literature-based radius value of the primary is derived from the photometric temperature of the primary for stars SAO~106989 and HD~205403. These radius values are further used to derive the secondary radius values based on the transit depth. Since the available data for both the stars SAO~106989 and HD~205403 is taken from ground-based photometry, the radii derived from photometric methods have significant differences than the values derived by us which are based on a detailed analysis using photometry data in unison with spectroscopy methods and Torres relation.  

%------------------------------------------------------------------------------------------------
%  TABLE 3
%------------------------------------------------------------------------------------------------

\begin{deluxetable*}{lcccccc}
\tiny
\tablecaption{Median values obtained from simultaneous RV and transit fitting for SAO~106989, HD~24465, EPIC~211682657 and HD~205403. It also includes data taken from literature for the respective sources; SAO~106989 by SW photometry as discussed in \cite{Street2007} (S07);
HD~24465 and EPIC~211682657 by K2 data as discussed in \cite{Barros2016} (B16) and HD~205403 by STEREO data \cite{Wraight2012} (W12). (The transit duration for HD~205403 was not mentioned in literature by \cite{Wraight2012}. Similarly, the information on radii of primary and secondary components of EBs, HD~24465 and EPIC~211682657, was not given in \citep{Barros2016}. These places are thus indicated by blanks in the table.)\label{tab:result_sao}}
\tabcolsep=0.11cm
\tablewidth{0.1pt}
\tablehead{
\colhead{Parameter} & \colhead{Units} & \colhead{SAO~106989} & \colhead{HD~24465} & \colhead{EPIC~211682657} & \colhead{HD~205403} & \colhead{Reference}
}
\startdata
Primary: &&&&&& \\
                 V mag. & & 	$9.3$ 		& $8.98$  & $8.69$ 		& $8.03$ 		& SIMBAD \\
		 Sp. Type & & 	F7 & 		F8 & 			  F2 & 			F2/F3			& SIMBAD \\
		 RA & WCS	&  $21h~16m~45.22s$ & $03h~54m~03.3689s$ & $08h~54m~33.0267s$ & $21h~35m~03.7303s$ & SIMBAD\\
		 Dec & WCS &  $+~19d~21m~36.79s$ & $+15d~08m~30.12s$ & $+15d~40m~55.030s$ & $-03d~44m~05.691s$ & SIMBAD \\ 
		 $M_{A}$ & M$_{\odot}$ & $1.11\pm0.22$\tablenotemark{a} & $1.337\pm0.008$\tablenotemark{a} & $1.721\pm0.047$\tablenotemark{a} & $1.445\pm0.019$\tablenotemark{a} & This work\\
		 $R_{A}$ & $R_{\odot}$ & $1.24$  & -- & -- & $1.46$ & S07,W12\\ 
		 $R_{A}$ & $R_{\odot}$ & $1.369\pm0.093$\tablenotemark{a} & $1.444\pm0.004$\tablenotemark{a} & $2.574\pm0.024$\tablenotemark{a} & $1.857\pm0.038$\tablenotemark{a} & This work\\
		 $\Omega_{1}$ & & $9.47\pm0.31$ & $5.988\pm0.010$ & $5.0155\pm0.113$  & $5.349\pm0.101$ & This work\\
		 ${\rm log}~g_{A}$& cgs & $4.211\pm0.127$\tablenotemark{a} & $4.245\pm0.0036$\tablenotemark{a} & $3.852\pm0.015$\tablenotemark{a} & $4.060\pm0.032$\tablenotemark{a} & This work\\
		 T$_{\rm{eff,A}}$  & K & $6000\pm100$ & $6250\pm100$ & $6650\pm150$ & $6600\pm100$ & This work\\
              	 $[$Fe$/$H$]$   & & $-0.2\pm0.1$ & $0.30\pm0.15$ &  $-0.1\pm0.15$ & $-0.100\pm0.15$ & This work\\			      
Secondary: &&&&&& \\
                 $e$& & $0.248\pm0.005$ & $0.208\pm0.002$ & $0.0097\pm0.0008$ & $0.002\pm0.002$ & This work\\
		 $\Omega_{2}$ & & $10.552\pm1.46$ & $14.663\pm0.056$ & $8.6954\pm0.0167$ & $7.299\pm0.686$ & This work \\
                 $\omega_*$& radians & $1.035\pm0.065$ & $5.988\pm0.010$ & $0.89\pm0.06$ & $5.603\pm0.165$ & This work\\
		 $P$& days& 4.400381 & 7.1977 & 3.141 & $2.4449\pm0.0005$ & S07, B16, W12\\
                 $P$& days & $4.39790\pm0.00001$ & $7.19635\pm0.00002$ & $3.142023\pm0.000003$ & $2.444949\pm0.000001$ & This work\\
	         $a sini$& AU & $0.0583\pm0.0005$ & $0.0849\pm0.0002$ & $0.0556\pm0.0005$ & $0.0438\pm0.0001$ & This work\\
	         $M_{B}$ & M$_\odot$ &  $0.256\pm0.005$\tablenotemark{a} & $0.233\pm0.002$\tablenotemark{a} & $0.599\pm0.017$\tablenotemark{a} & $0.406\pm0.005$\tablenotemark{a} & This work\\
		 $R_{B}$ & R$_\odot$ & $0.123$  & -- & -- & $\geqslant0.35$ & S07, W12 \\
	         $R_{B}$ & R$_\odot$ & $0.326\pm0.012$\tablenotemark{a} & $0.244\pm0.001$\tablenotemark{a} & $0.566\pm0.005$\tablenotemark{a} & $0.444\pm0.014$\tablenotemark{a} & This work\\
		 T$_{\rm{eff,B}}$ & K & $2380.28\pm259.39$ & $2335.6\pm8.56$ & $4329.0\pm49.42$ & $4651\pm123.33$ & This work\\
		 ${\rm log}~g_{B}$& cgs & $4.818\pm0.128$ & $5.029\pm0.007$ & $4.711\pm0.015$ & $4.752\pm0.033$ & This work\\
RV: &&&&&&\\
		$K$& km~s$^{-1}$ & $26.189\pm0.251$\tablenotemark{a} & $18.629\pm0.053$\tablenotemark{a} & $49.691\pm0.636$\tablenotemark{a} & $42.7785\pm0.2627$\tablenotemark {a} & This work\\
                $M_{B}/M_{A}$ && $0.230\pm0.002$ & $0.174\pm0.008$ & $0.3481\pm0.004$ & $0.2861\pm0.0053$ & This work\\
		$M sin^3i $ & M$_\odot$ & $1.324\pm0.027$\tablenotemark{a} & $1.560\pm001$\tablenotemark{a} &$2.312\pm0.063$\tablenotemark{a} & $1.7986\pm0.0234$\tablenotemark{a} & This work\\
                $\gamma$ &  km~s$^{-1}$ & $2.801\pm0.154$ & $-15.800\pm0.029$ & $28.629\pm0.336$ & $14.745\pm0.132$ & This work\\
		$Age$\tablenotemark{b} &  \textbf{Gyr} & $\sim~2$ & $\sim~2.3$ & $\sim~1.4$ & $\sim~1.2$ & This work\\		
Transit: &&&&&&\\
		$T_C$ & BJD &  $2456595.968\pm0.028$ & $2457097.860675\pm0.004145$ & $2457880.386430\pm0.009466$ & $2457878.9245\pm0.0325$ & This work \\
		$i$ & degrees & $81.624\pm0.547$ & $86.267\pm0.013$ & $87.113\pm0.037$ & $82.103\pm0.146$ & This work \\
		$\delta$ & mag & $0.0135$ & $0.038\pm0.002$ & $0.050\pm0.0006$ & $0.057\pm0.014$ & S07, B16, W12\\
		$\delta$ & mag & $0.063\pm0.002$ & $0.0315\pm0.0005$ & $0.053\pm0.001$ & $0.063\pm0.003$ & This work \\
		$T_{14}$ & min & $145$ & $263\pm14$ & $375\pm5$ & $-$ & S07, B16 \\
\enddata
\tablenotetext{a}{uncertainities estimated using the error propagation.}
\tablenotetext{b}{Average stellar age determined isochronically and gyrochronically. See Sec~\ref{sec:discussion}}
\end{deluxetable*}

%------------------------------------------------------------------------------------------------

\subsubsection{SAO~106989} 
The periodicity for SAO~106989 of $4.39790\pm0.00001$~d obtained from the analysis is close to the value obtained from SW photometry. The RV semi-amplitude for the EB system is $26.189\pm0.251$~km~s$^{-1}$ with an eccentricity of $0.248\pm0.005$ at an orbital separation of $0.0583\pm0.0005$ AU. The top left panel of Fig.~\ref{fig:rv_sao} illustrates the RV versus orbital phase for SAO~106989. Solid red circles (top sub-figure) show RV measurements of the star taken with PARAS. The figure also shows the residuals (Observed-Model) in the bottom sub-figure of the panel. \textit{PARAS SPEC} routine applied on SAO~106989 gives $T_{\rm eff}$~=~$5963~\pm~100$~K, $[Fe/H]$~=~-$0.2~\pm~0.1$ and ${\rm log}~g$~=~$4.23~\pm~0.1$ (The \textit{PARAS SPEC} results for all the sources studied here are mean of the results obtained from synthetic spectral fitting and equivalent width method). The mass and radius for the primary of the EB system, SAO~106989, based on the spectroscopic analysis and Torres relation \citep{Torres2010} are $1.111\pm0.27$~M$_\odot$ and $1.369\pm0.111$ R$_\odot$ respectively. The mass of the secondary derived from RV data is $0.256\pm0.005$~M$_\odot$ determined at an accuracy of $\sim~3$~per~cent (formal errors). The radius value predicted for SAO~106989B from SW photometry is $R_{B}=0.126 R_{\odot}$. This is much lower than theoretically expected radius value derived for a star having a mass of $M_{B}=0.256$~M$_\odot$. However, we retrieved KELT data and performed detailed transit modeling. Despite the data being slightly noisy, we could simultaneously fit the transit obtained from KELT lightcurves. The top left panel of Fig~\ref{fig:trans_sao} (upper sub-figure) shows the simultaneous fit for the transit light curve obtained by using KELT light curve (filled circles) overplotted with the model derived from PHOEBE (solid curve) with residuals being plotted in the lower sub-figure of the panel. The simultaneous fit gives us a transit depth of $0.063\pm0.002$ mag and angle of inclination of $81.624\pm0.547^{\circ}$. The radius determined through observations is $0.326\pm0.012~R_{\odot}$. 
%Though the error bars are larger due to noisy photometry data, the radius measurements are consistent with the theoretical predictions within the error bars.

\subsubsection{HD~24465} 
EB HD~24465 is a short period EB candidate by K2 photometry having a transit depth of 38~mmag \citep{Barros2016}. We confirmed the orbital period of this EB at $7.19635\pm0.00002$~d with PARAS RV data. The RV semi-amplitude for the EB system is $18.629\pm0.053$~km~s$^{-1}$ with an eccentricity of $0.208\pm0.002$ at an orbital separation of $0.0849\pm0.0002$ AU. The top right panel of Fig~\ref{fig:rv_sao} illustrates the RV versus orbital phase for HD~24465. \textit{PARAS SPEC} routine applied on HD~24465 gives $T_{\rm eff}$~=~$6200_{-81}^{+76}$, $[Fe/H]$~=~$0.30\pm0.14$ and ${\rm log}~g$~=~$4.03\pm0.15$. Based on stellar parameters derived and application of Torres relation \citep{Torres2010}, the mass and radius for HD~24465A are derived as $1.337\pm0.008$~M$_\odot$ and $1.444\pm0.003$ R$_\odot$ respectively. The top right panel of Fig~\ref{fig:trans_sao} (upper sub-figure) shows the simultaneous fit for the transit light curve obtained by using K2 data (filled circles) overplotted with the model derived from PHOEBE (solid curve) with the residuals plotted in the lower sub-figure of the panel. The simultaneous fit gives us a transit depth of $0.03145\pm0.0005$ mag and angle of inclination of $86.267\pm0.013^{\circ}$. The mass and radius of the secondary derived here are $0.233\pm0.002$~M$_\odot$ and $0.244\pm0.001$~R$_\odot$ determined at an accuracy of $\sim~1$~per~cent (formal errors). 
%This value is in agreement with the mass-radius relation \citep{Baraffe1998}.

\subsubsection{EPIC~211682657} 
EPIC~211682657 is an EB with a periodicity of $3.142023\pm0.000003$~d reported by K2 photometry, which was confirmed by us with the RV data. The RV semi-amplitude for the EB system is $49.691\pm0.636$~km~s$^{-1}$ with a small eccentricity of $0.0097\pm0.0008$ . The EB has an orbital separation of $0.0556\pm0.0005$ AU. The bottom left panel of Fig~\ref{fig:rv_sao} illustrates the RV versus orbital phase for EPIC~211682657. \textit{PARAS SPEC} routine applied on EPIC~211682657 gives $T_{\rm eff}$~=~$6650\pm125$, $[Fe/H]$~=~$-0.1\pm0.15$ and ${\rm log}~g$~=~$3.8\pm0.15$. The mass and radius derived for the primary star of the EB, EPIC~211682657, are $1.721\pm0.048$~M$_\odot$ and $2.574\pm0.024$~R$_\odot$ respectively \citep{Torres2010}. The bottom left panel of Fig~\ref{fig:trans_sao} (upper sub-figure) shows the simultaneous fit for the transit light curve obtained by using K2 data (filled circles) overplotted with the model derived from PHOEBE (solid curve). The residuals are plotted in lower sub-figure of the panel. We determine a transit depth of $0.053\pm0.0008$ mag and angle of inclination of $87.113\pm0.0368^{\circ}$ from the simultaneous fit. The mass and radius of EPIC~211682657B based on the combined fit are $0.599\pm0.017$~M$_\odot$ and $0.566\pm0.005$~R$_\odot$ respectively derived at an accuracy of $\sim~1$~per~cent (formal errors). 
%The radius value is higher than those expected from theory by $13\%$ \citep{Baraffe1998}.

\subsubsection{HD~205403}
HD~205403 is another short period EB with a periodicity of $2.444949\pm0.000001$~d as mentioned by \cite{Wraight2012} from STEREO photometry. We retrieved the STEREO archival data and confirmed the periodicity with the transit data as well as the RV data of PARAS. The RV semi-amplitude for the primary of HD~205403 EB system is $42.7785\pm0.2627$~km~s$^{-1}$ with a near circular orbit having eccentricity of $0.002\pm0.002$. The two stars of the EB are separated by $0.0438\pm0.0001$ AU. The bottom right panel of Fig~\ref{fig:rv_sao} illustrates the RV plotted against orbital phase for HD~205403. \textit{PARAS SPEC} routine applied on HD~205403 gives $T_{\rm eff}$~=~$6525\pm100$, $[Fe/H]$~=~$0.1\pm0.14$ and ${\rm log}~g$~=~$3.6\pm0.15$. Torres relation \citep{Torres2010} when applied to HD~205403A gives us the mass and radius as $1.445\pm0.089$~M$_\odot$ and $1.857\pm0.038$~R$_\odot$ respectively. The bottom right panel of Fig~\ref{fig:trans_sao} (upper sub-figure) shows the simultaneous fit for the transit light curve obtained by using K2 data (filled circles) overplotted with the model derived from PHOEBE (solid curve) with residuals plotted in the lower sub-figure of the panel. The RV and photometry data is fitted simultaneously giving us a transit depth of $0.063\pm0.0027$ mag and angle of inclination of $82.103\pm0.146^{\circ}$. We determine the mass and radius of the secondary as $0.406\pm0.005$~M$_\odot$ and $0.444\pm0.014$~R$_\odot$ respectively. The accuracy for determination of mass and radii is $\sim~6$~per~cent (formal errors). 
%The radius value is higher than those expected from theory by $15\%$ \citep{Baraffe1998}.
  
%****************************************************************************************************************

\section{Discussion} \label{sec:discussion}

\subsection{Tidal evolution in EBs}

The primary stars for all the EBs, SAO~106989, HD~24465, EPIC~211682657 and HD~205403 are F type primaries. F type stars act as bridge between solar type stars having large convective envelopes and early type stars having radiative envelopes. Stars having larger convective zones suffer faster tidal dissipation than those having outer radiative envelopes \citep{Zahn1977}. Turbulent friction acting on the equilibrium tide acts on the convective zones whereas radiative damping of the dynamical tide on the radiative zones are the chief progenitors for tidal disspipation \citep{Zahn2008}. The nature of tidal interaction depends more on separation of the two components rather than their sizes \citep{Ogilvie2014}. The tidal forces work in the direction to synchronize spin and angular velocities through an exchange of angular momentum and the dissipation of energy, alignment of spin axis perpendicular to orbital plane and circularization of the binary orbit \citep{Mathis2009}. Tides caused by close-in companions pose threat to the existence of the binary system in few cases. If the spin of the primary star is slower than the binary orbital period, tidal torque raised by the companion will spin up the primary. In order to conserve angular momentum, semi-major axis of companion will decrease resulting to an inward spiralling of the companion towards the primary. This happens to G and K type primaries whereas for F type primaries the spin period is sufficiently high to evade this engulfment \citep{Poppenhaeger2017, Bouchy2011, Bouchy2011b}. This is the main reason we see F+M systems commonly in nature. 

The rotational velocity (v~sin~i) of SAO~106989 is $\sim 20$ km~s$^{-1}$ as computed from the RV cross-corelation function (CCF). We assumed here that the primary star's rotation axis is aligned with the orbital inclination. Thus, this is the minimum rotational velocity inferred for the star and thus the rotational period derived from here will be maximum. In Fig~1 of \cite{Meibom2015}, the authors have compared the rotational periods, temperatures and ages of stars. We use the rotational period of SAO~106989 to estimate the age of the EB to be between $0.7-1$~Gyr. The second source, HD~24465 has a higher temperature than SAO~106989 but has a smaller rotational velocity of $\sim 11$ km~s$^{-1}$ as computed from the CCF width. We thereby estimate an age of $\sim~2$~Gyr on account of the rotational period of the star. This age is more than that of SAO~106989 and thereby we conclude that HD~24465 has slowed down based on its age. The next source, EPIC~211682657 is an early F type star, having higher temperature than the other two stars discussed. It has a large rotational velocity of $\sim~40$ km~s$^{-1}$ as computed from CCF width. We, similarly derive the age of $\sim~1.0$~Gyr for this EB based on the rotational period of the primary star. Finally, HD~205403 is also a mid F type star like EPIC~211682657. It has a temperature close to 6500~K. It has a rotational velocity of $\sim~25$ km~s$^{-1}$ and thereby we derive an age of $\sim~1.0$~Gyr for this EB. We also utilized the publically available ISOCHRONES package {\citep{Morton2015} to determine the age of these EB systems. We used Dartmouth Stellar evolution Tracks \citep{Dotter2008} for the models and provided the stellar parameters (for the primary star), $T_{\rm{eff}}$, $[Fe/H]$, ${\rm log}~g$ derived from \S~\ref{sec:spec_analysis}. The photometric magnitudes in different bands (B, V, J, H, K) were taken from SIMBAD. The ages derived from these ischrones are $3.047\pm0.85$ and $2.517\pm0.45$ Gyr for SAO~106989 and HD~24465 respectively. Similarly, the ages of EPIC~211682657 and HD~205403 are $1.705\pm0.393$ and $1.44\pm0.207$ respectively. The ages inferred from the rotational periods of the EBs and those derived by the Dartmouth Stellar evolution tracks more or less agree for all the EBs except for SAO~106989. The age of SAO~106989 derived from its rotational period is almost three times shorter than that derived from the ISOCHRONES package. However, for stars in close binary systems the tides generated by the M dwarf companion may spin up the primary star, SAO~106989A. Thus, the rotational velocity of SAO~106989A would be larger and thereby its rotational period is smaller as compared to had the star been isolated. The ages of the systems that we used for further analysis are the average of those derived by the above two methods, as indicated in Table~\ref{tab:result_sao}.

Synchronization of orbital and rotational velocities is an indication of stable evolution of the orbit of the system \citep{Hut1981}. Several of binaries are studied for their synchronization and circularization timescales \citep{Meibom2006, Claret1995}. For stars with convective envelopes (mass $\leq$~1.6~M$_\odot$) and solar ages, \cite{Zahn1977} assumed that the primary star rotates uniformly with an angular velocity $\omega$ and its spin axis is perpendicular to the orbital plane in a similar reference rame corotating with the star. The authors derived the synchronization timescales in years as given by eqn. $t_{sync} \sim q^{-2}(a/R)^6 \sim 10^4 ((1+q)/2q)^2 P^4 $. Here, q is the mass ratio of the two stars, a is the orbital separation, R is the radius of the primary star and P is the orbital period of the system. We used the above-mentioned parameters needed in this equation from Table~\ref{tab:result_sao} and thereby estimate the synchronization time scale for SAO~106989 to be $\sim$~2~Myr. Since, the age of the star is more than this, we rightly see the orbital and rotational velocity for the star synchronized with each other. For HD~24465, we similarly estimate the synchronization time scale to be $\sim$~28~Myr. Here too, the age of the star is more than the synchronization timescale, and we see the orbital and rotational velocities synchronized in this case as well. For EPIC~211682657, the synchronization time scale is very small, $\sim$~0.03~Myr, due to its large mass ratio (q) and small period and the same for HD~205403 EB is 0.2~Myr. These synchronization timescale values are similarly larger than the respective ages of the two EBs and thereby we infer that all EBs are synchronized.

The circularization time scale as mentioned in \cite{Zahn1977}, is given is years as $t_{circ} \sim (q(1+q)/2)^{-1}~(a/R)^8 \sim 10^6 q^{-1} ((1+q)/2)^{5/3} P^{16/3}$. For SAO~106989, this value is $\sim$~5~Gyr. The same value for HD~24465, EPIC~211682657, HD~205403 are $\sim$~88~Gyr, $\sim$~0.67~Gyr and $\sim$~0.2~Gyr respectively. It is important to note that these estimations are based on the assumption that these stars have a convective envelope. Recently, \cite{Eylen2016} studied the orbital circularization rates of hot and cool stars from the Kepler EB catalog. The authors found that EBs having both components as hot-hot type ($\geq~6250$~K) are more probable to have eccentric systems as compared to EBs having cool-cool ($\leq~6250$~K) and a combination of hot-cool systems. This is mainly due to the tidal efficiency rate, which is dependent on the total mass and orbital period of the EB. \cite{Zahn1977} derives a lower limit on R$_{*}$/a (inverse of relative separation) as 0.025 for synchronization. Systems below this relative separation are found to be non-synchronized. It is also important to note that these trends of R$_{*}$/a and eccentricity are for solar age and composition. Orbits for systems having a lower limit of R$_{*}$/a~$\sim~0.25$ are circular. Systems having a relative radius value smaller than 0.25 are eccentric in nature. Thus, we see circularization is a much slower process than synchronization. From the current work, R$_{*}$/a for SAO~106989 is 0.11 and that for HD~24465 is 0.08. The relative radii for both these systems are larger than 0.025 but very small as compared to 0.25. Both the EBs have eccentricity greater than 0.2. Thus, we rightly conclude that the circularization timescales for these EBs are more than their ages. Though these EBs are synchronized for their rotational and orbital periods, they have not yet circularized. For the other two EBs, EPIC~211682657 and HD~205403, the R$_{*}$/a for both these EBs is 0.21 and 0.19 respectively, which are relatively larger values than those for SAO~106989 and HD~24465. R$_{*}$/a values are sufficiently larger than the synchronozation limit and is also comparable to the circularization limit. Moreover, the derived circularization timescales are comparable to the respective ages of the EBs. Thereby, we see these EBs are not only synchronized for their rotational and orbital periods but also have negligible eccentricities as compared to the other two EBs.

In order to analyze this argument carefully, we have compiled all the F+M systems characterized for their masses and orbital parameters from literature. Out of the 97 F+M EBs, a major set of samples (75) come from the recent paper \cite{Triaud2017} and the remaining 22 sources are from \cite{Von2017, Eigmuller2016, Chaturvedi2014, Zhou2014, Tal-Or2013, Ofir2012, Fernandez2009, Beatty2007, Pont2006, Bouchy2005}. In Fig~\ref{fig:ecc}, we have plotted eccentricity vs Period for these 97 F+M EBs in left panel and eccentricity vs secondary mass (M$_{2}$) in right panel. The error bars (not shown in the scatter plots) on eccentricity and M$_{2}$ are on an average between $2-5\%$ of the actual values. We have overplotted the F+M EBs studied as a part of this work on the eccentricity vs Period and eccentricity vs M$_{2}$ plots in red filled triangles in Fig~\ref{fig:ecc} in left and right panels respectively. As expected, we see that F+M EBs having short orbital periods are mostly circular and the ones having longer periods show range of eccentricities. The scatter seen in EB parameters can be attributed to different methods adopted for analyses. This is consistent with the tidal circularization theory by \cite{Zahn1977}. As seen from the right panel of the plot, less massive secondary companions have range of eccentricities and as the mass of the companion increases, the systems tend to show more circular orbits. The mass ratio, q, affects the tidal circularization rate. However, it is also important to note that the observed eccentricities is a function of initial eccentricity at the time of system formation and thereby larger the primordial eccentricity, circularization timescales would be larger \citep{Mazeh2008}. 

Two of the EBs studied here, SAO~106989 and HD~24465 follow the trends marginally with large eccentricities despite their short orbital periods. These EBs are thereby unique as they belong to a category of handful of such systems. EPIC~211682657 and HD~205403 are close to circular  (e~$\sim0.009$ and $\sim0.002$ respectively). The primaries for these two EBs are mid-F type stars with higher rotational velocities. The tidal dissipation rates in such sytems are expected to be lesser than for the other two EBs. Short period eccentric EBs have higher probability of hosting a distant third body as a perturber \citep{Mazeh2008}. Long term monitoring on these targets shall help discover any such trends if present. Moreover, such systems are also prone to show wide range of obliquities for spin-orbit orientation as compared to stars with convective exteriors \citep{Winn2010}. Thus, a detailed investigation of SAO~106989 and HD~24465 on a longer timeline will be desirable in future.

%------------------------------------------------------------------------------------------------
%  FIG 5
%------------------------------------------------------------------------------------------------

\begin{figure}
\centering	
\includegraphics[width=0.48\textwidth]{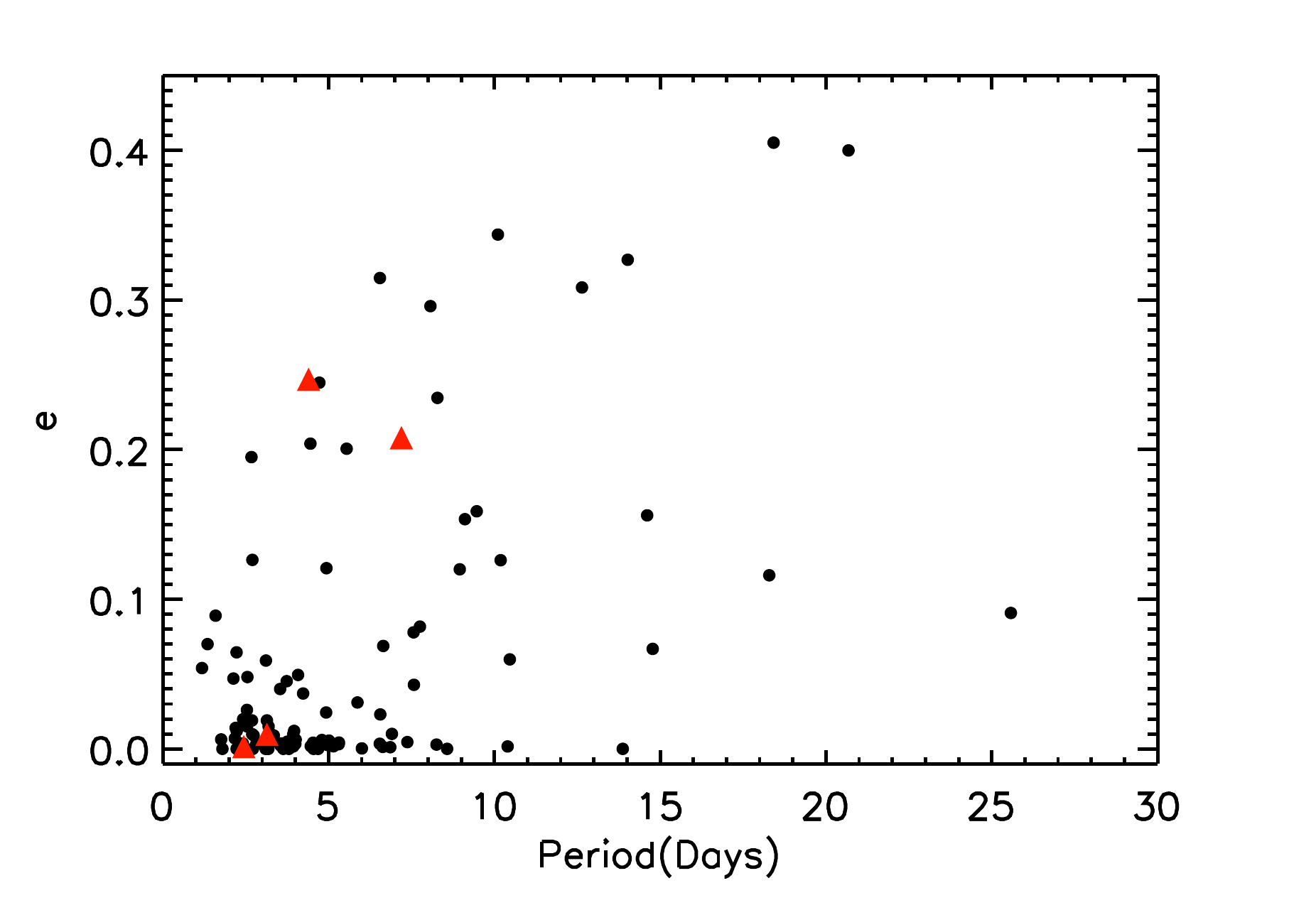}
\includegraphics[width=0.48\textwidth]{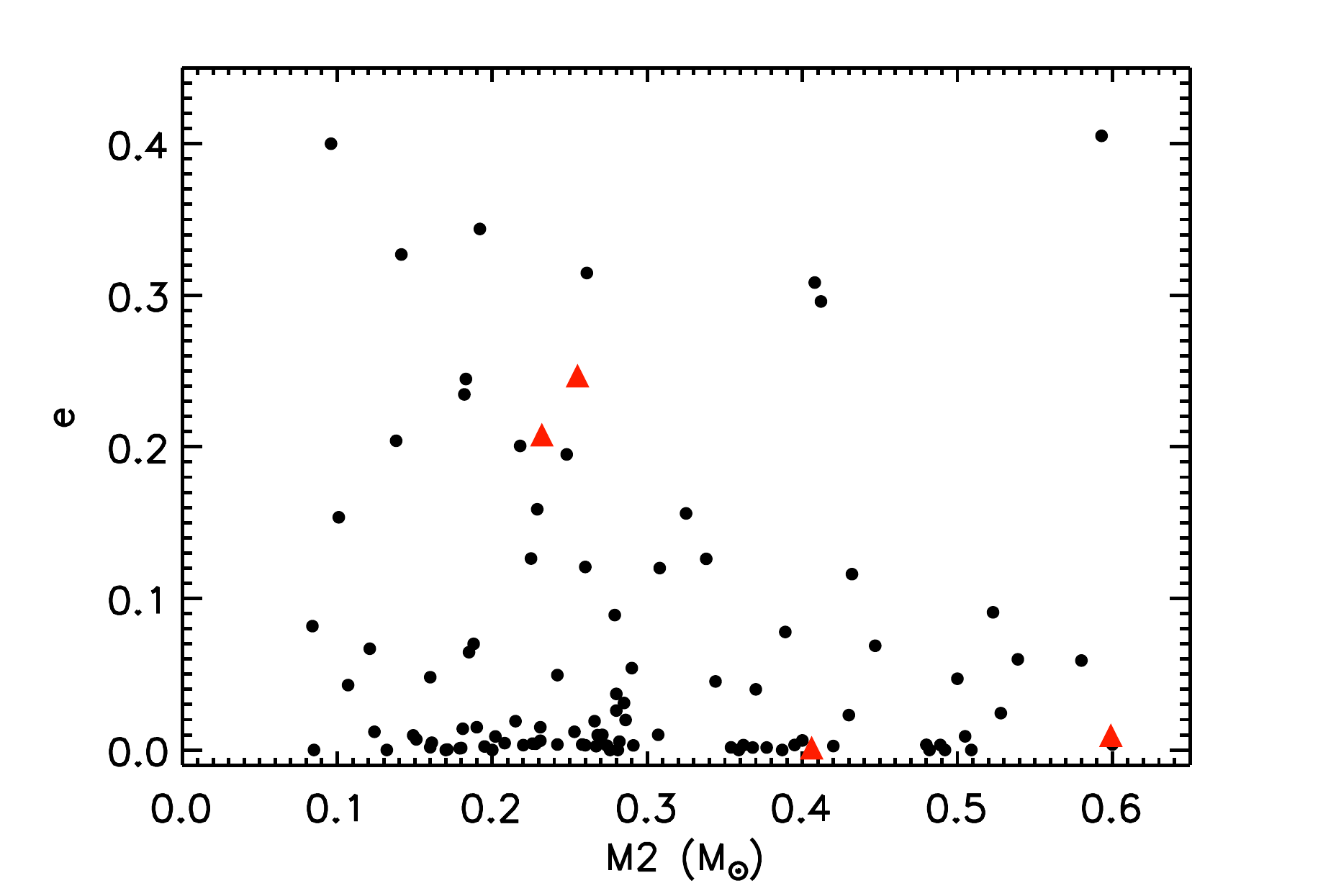}
\caption{(Left panel) Scatter plot for Eccentricity vs Period for the 97 F+M Ebs compiled from literature in black open circles. Overplotted are the four EBs studied as a part of this work in red filled triangles.
(Right panel) Scatter plot for Eccentricity vs Companion mass (M$_{2}$) for the 97 F+M EBs compiled from literature in black open circles. Overplotted are the four EBs studied as a part of this work in red filled triangles. The literature sources are taken from \cite{Triaud2017, Von2017, Eigmuller2016, Chaturvedi2014, Zhou2014, Tal-Or2013, Ofir2012, Fernandez2009, Beatty2007, Pont2006, Bouchy2005, Pont2005a}} \label{fig:ecc}
\end{figure}

%------------------------------------------------------------------------------------------------

\subsection{Mass-Radius relation} \label{subsec:mr}

We have inferred the radii of the program stars using photometry observations and the empirical Torres relation. A comparison for testing the isochrone-predicted M dwarf parameters against the Torres relation is in order. We compared the derived radii of the low mass companions in the EB systems based on our current work with Baraffe's grid of new models \citep{Baraffe2015} that have updated molecular linelists, revised solar abundances and the line opacities for several important molecules. These updated models have been able to account for some of the flaws of the previous Baraffe's models \citep{Baraffe1998} such as predicting optical colours of the stars that are too blue \citep{Baraffe2015}. The masses of the M dwarfs, detected as companions to F type stars discussed in this paper, range from 0.232 to 0.599~M$_{\odot}$.

The age of all the primary stars of the EBs are between 1-3 Gyr as discussed in the previous section. From our RV analysis, we find that SAO~106989B has a mass of $0.256\pm0.005$M$_{\odot}$. The \cite{Baraffe2015} models for 1 Gyr isochrones, the radius for $0.25$~M$_{\odot}$ turns out to be $0.26$~R$_{\odot}$ for [M/H]~=~0.0. The value retrieved from fitting of KELT light curve is $R_{B}=0.326\pm0.012$~R$_{\odot}$. The similar estimate given by SW photometry is $R_{B}=0.126$~R$_{\odot}$. The noisy SW photometry data may have lead to a diluted measurement of transit depth. Though the larger error bars cannot be ignored for the derived values it is worth mentioning that the observationally derived radius for SAO~106989B is $20\%$ larger than the theoretically derived values. 
From our current study, we have derived the mass of HD~24465B as $0.233\pm0.002$~M$_\odot$. For [M/H]~=~0.0, a $0.233~M_{\odot}$ star has a radius of $0.23~R_{\odot}$ \citep{Baraffe2015}. The value derived from K2 photometry matches within error bars of the predicted model. We have derived the mass of EPIC~211682657B as $0.599\pm0.017$~M$_\odot$ based on the RV data from PARAS. We derive a radius of $0.566\pm0.005$ from K2 photometry. The same value derived theoretically from \cite{Baraffe2015} models is $\sim$$~0.557~R_{\odot}$ for [M/H]~=~0.0. The observed value is matches the theoretically derived radius value. The mass derived for HD~205403B is $0.406\pm0.005$~M$_\odot$ from our current RV data. The value for radius derived from STEREO photometry is $0.444\pm0.014$. We derive the theoretical value for the radius as $\sim$$~0.37~R_{\odot}$ from \cite{Baraffe2015} models, which is $17\%$ lesser than the observed radius value.

In Fig~\ref{fig:vlms} is plotted a Baraffe isochrone for 1 Gyr and solar metallicity \citep{Baraffe2015} in the Mass-Radius space (black solid line). Over-plotted on this diagram are the observationally derived values for the four stars studied as a part of this work (red filled triangles). Also shown are the results taken from literature for M dwarfs (M~$\leq0.6M_{\odot}$) which have masses and radii measured at best upto $10\%$ (See Table~\ref{tab:known_ebs} in \S~\ref{sec:app} for the sources taken from literature). From the figure, we see a disagreement between the observed radii of the stars and its theoretical predictions beyond 0.3~M$_{\odot}$. M dwarfs below this mass limit, seem to follow the theoretical M-R relation within error bars. Above this mass limit, we see a huge scatter, which points towards a higher observationally derived radius value. Two of the stars as a part of this work follow a similar trend as that of the stars seen in literature. The larger error bars on SAO~106989B are due to the relatively noisy KELT dataset. The case is similar with HD~205403B, which has data from STEREO photometry. Both these stars have radii $17-20\%$ larger than the theoretically predicted values. The remaining two stars HD~24465B, and EPIC~211682657B have observed radii consistent with predictions from theory. 0.3~M$_{\odot}$ is the mass limit between stars that are fully convective and the ones that have radiative cores. Convection is the most efficient mechanism of energy transport in low mass regime. The central density for the stars, which are fully convective (below $0.3~M_{\odot}$), decreases with the hydrogen burning phase. With reduced central densities, electron degeneracy effects dominate in stellar interior affecting thermal efficiency and further inhibiting flux transport. This inhibition leads to increase in the stellar radii \citep{Cassisi2011}. Strong magnetic fields inhibit convection causing inflation of stellar radii \citep{Lopez-Morales2005, Mullan2001}. Single stars are known to be slow rotators whereas many of the binaries are fast rotators depicting strong indications of X-ray activity from the corona and the H$_{\alpha}$ activity from the chromosphere \citep{Chabrier2007}. Thus, the magnetic activity level for binaries can be 100 times more than the single stars \citep{Mullan2001}. Another possible scenario causing mismatch in the observationally computed and theoretically derived radius are starspots seen as dark regions seen on the observable photosphere of the star due to the presence of local magnetic fields which suppress the convective motion and thereby energy transport from the stellar interior to the surface \citep{Strassmeier200​9​}. \cite{Chabrier2007} concluded in their study that the inhibition of convection in fast rotating stars and the presence of stars spots on the stellar disk could affect the stellar models. Cool starspots too are reflective of the inhibition of energy by convective transport in the interior of the star. There is a possibility that the scatter in observational radii could be due to the large range of metallicities and stellar activity of the samples \citep{Lopez-Morales2007}. \cite{Berger2006} in their study find that the disagreement is larger among metal-rich stars than metal-poor stars. They conclude that current atmospheric models have missed some opacity components which may lead to a larger radii for stars having higher metallicity. If we consider stars having same mass, a decrease in stellar metallicity leads to decrease in opacity. This in turn causes raised electron degeneracy leading to inflated stellar radii \citep{Cassisi2011}. An improper modelling of the molecular absorption coefficients due to incorrect abundance analysis results in an erroneous M-R relationships \citep{Berger2006}. \cite{Lopez-Morales2007} showed that stars with $[Fe/H] >-0.25$ show larger deviations in the radius measurements from the models than stars with $[Fe/H] <-0.25$. However, this issue needs to be further investigated. Therefore, it becomes imperative to detect and study more such systems and determine their masses and radii to very high precision.

%------------------------------------------------------------------------------------------------
%  FIG 5
%------------------------------------------------------------------------------------------------

\begin{figure}
\centering	
\includegraphics[width=0.65\textwidth]{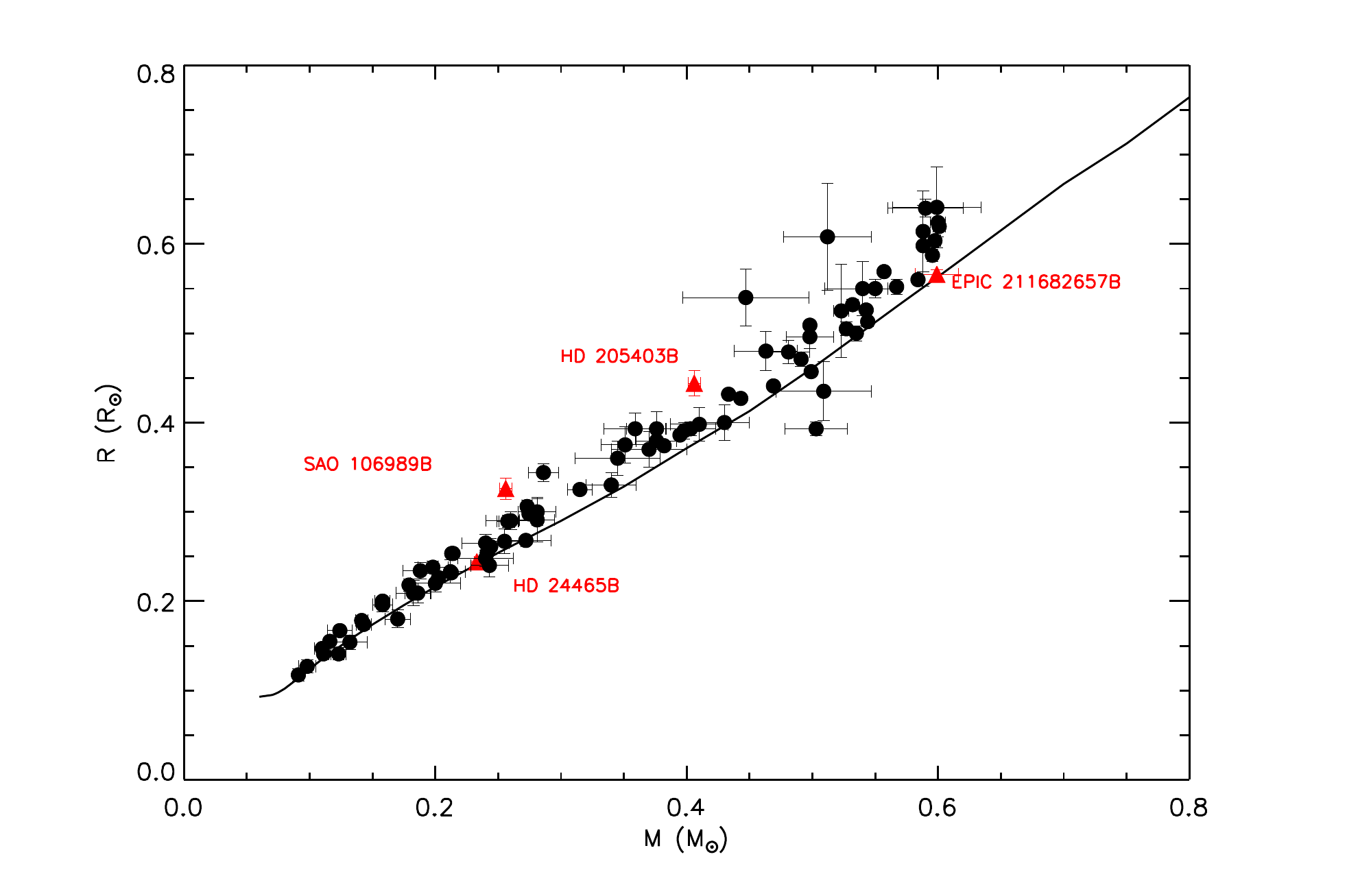}
\caption{Mass-Radius diagram for M dwarfs based on Baraffe models for 1 Gyr isochrone and solar metallicity. Overplotted in black filled circles are the M dwarfs taken from literature as shown in Table~\ref{tab:known_ebs} and the ones in red filled triangles are studied as a part of this paper. The masses and radii are plotted with their respective error bars.} \label{fig:vlms}
\end{figure}

%------------------------------------------------------------------------------------------------

Future spectroscopic observations and detailed photometry for all these stars during their respective transits may enable us to observe the Rossiter-McLaughlin (RM) effect \citep{Gaudi2007}, and help determine whether the secondary star is in retrograde or prograde orbital motion with respect to the rotation of the primary. This may lead to a better understanding of the binary formation mechanisms at a primordial stage. M dwarfs peak more in the near infra-red and we expect the spectra of the secondary to be seen with a larger telescope, the case of SB$_{2}$ systems. Since the companions are M dwarfs, future high resolution near-infrared observations with instruments sensitive in the infrared wavelength region like the upcoming HPF \citep{Mahadevan2014} and CARMENES \citep{Quirrenbach2010} will be able to provide accurate masses and radii of the companion M dwarfs.
 
%Ground-based observations will be desirable in future to assess the radius of the components and inclination for the system at high accuracies.

\section{Summary} \label{sec:summary}

We have detected and characterised four F+M EBs, SAO~106989, HD~24465, EPIC~211682657 and HD~205403, in short orbital periods from SuperWasp, STEREO and K2 EB candidate databases using RV data from PARAS and light-curve data from the respective photometry archives for the stars. The prominent results are summarized below:
\begin{itemize} [noitemsep,topsep=0pt]
\item Masses for the companion M dwarfs are determined as $0.256\pm0.005$~M$_\odot$, $0.233\pm0.002$, $0.599\pm0.017$ and $0.406\pm0.005$~M$_\odot$ respectively. 
\item The radii for the M dwarf companions are found to be $0.326\pm0.012$, $0.244\pm0.001$, $0.566\pm0.005$ and $0.444\pm0.014$~R$_\odot$ respectively. Since the error bars on radius measurements for SAO~106989 and HD~205403 are relatively larger, precision photometry measurements in future are desirable. 
\item One of these M dwarfs, HD~24465, with mass less than $0.3~M_{\odot}$ is found to have radius which is in good agreement with the theoretical predictions whereas the other one observed with KELT, SAO~106989 shows discrepancies mostly attributed to noisy data. The radius for the EB HD~205403 having mass greater than $0.3~M_{\odot}$ has $17\%$ higher value than the theoretically derived ones whereas for the case of EPIC~211682657 is consistent with theory. Stars less massive than $0.3~M_{\odot}$ have totally convective interiors and are thus believed to follow the theoretical M-R relation. 
\item We have estimated the rotational and orbital velocities for these EBs and found them to be synchronized as expected theoretically. Out of the four EBs, SAO~106989 and HD~24465 show significant eccentricities whereas EBs, EPIC~211682657 and HD~205403 have smaller eccentricities.
\end{itemize}

Future long-term follow-up for these systems is essential. Similar studies of EBs in near future will help clarify observational biases associated with the stellar evolutionary models.

\textbf{Acknowledgments:}
PARAS spectrograph is fully funded and being supported by Physical Research Laboratory (PRL), which is part of Department of Space, Government of India. The authors would like to thank Director, PRL for his support. The data pipeline development was done in collaboration with Suvrath Mahadevan and Arpita Roy at Pennsylvania University, USA. We acknowledge the help from Vaibhav Dixit, Vishal Shah, Arvind Rajpurohit and Mount Abu Observatory staff for their support during observations. This research has made use of the ADS and CDS databases, operated at the CDS, Strasbourg, France.}

\section{Appendix}
\label{sec:app}
%------------------------------------------------------------------------------------------------
%  TABLE 4
%------------------------------------------------------------------------------------------------
\startlongtable
\tiny
\begin{deluxetable*}{lccc}
%\begin{tiny}
\label{tab:known_ebs}
\tablewidth{0pt}
\tablecaption{A compilation of known M dwarfs (M~$\leq0.6M_{\odot}$) from literature other for masses and radii measured at accuracies better than or at best equal to 10$\%$. M dwarfs studied in this paper are indicated in bold font.}
\tablehead{
\colhead{Name of EB} & \colhead{Mass} & \colhead{Radius} & \colhead{Reference} 
}
\startdata
J1219-39B 		&   $0.091\pm0.002$ 		&   $0.1174\pm0.0071$ 		& (1) \\
J2343$+$29$^{\star}$ 		&   $0.098\pm0.007$		&   $0.127\pm0.007$  		& (2) \\
HATS550-016B		&   $0.110\pm0.006$		&   $0.147\pm0.004$ 		& (3) \\
NNSer-B			&   $0.111\pm0.004$		&   $0.141\pm0.002$ 		& (4) \\
GKVir			&   $0.116\pm0.003$		&   $0.155\pm0.003$ 		& (5) \\
GJ551			&   $0.123\pm0.006$		&   $0.141\pm0.007$ 		& (6) \\
HAT-TR-205		&   $0.124\pm0.010$ 		&   $0.167\pm0.006$ 		& (7) \\
HATS551-021B		&   $0.132\pm0.014$		&   $0.154\pm0.008$ 		& (3) \\
KIC1571511B		&   $0.14136\pm0.0051$		&   $0.17831\pm0.0013$ 		& (8) \\
WTS19g4-020B		&   $0.143\pm0.006$		&   $0.174\pm0.006$ 		& (9) \\
GJ699			&   $0.158\pm0.008$		&   $0.196\pm0.008$ 		& (6) \\
SDSSJ1210$+$3347		&   $0.158\pm0.006$		&   $0.20\pm0.003$ 		& (10) \\
HATS551-019B		&   $0.17\pm0.01$		&   $0.18\pm0.01$ 		& (3) \\
HATS551-027B		&   $0.179\pm0.002$		&   $0.218\pm0.007$ 		& (11) \\
RRCaeB			&   $0.1825\pm0.0139$		&   $0.209\pm0.0143$		& (12) \\
J0113+31B		&   $0.186\pm0.010$		&   $0.209\pm0.011$ 		& (13) \\
2MASS02405152$+$5245066	&   $0.188\pm0.014$		&   $0.234\pm0.009$ 		& (14) \\
T-Lyr1-01662		&   $0.198\pm0.012$		&   $0.238\pm0.007$ 		& (15) \\
HATS553-001B		&   $0.20\pm0.02$		&   $0.22\pm0.01$ 		& (3) \\
KEPLER16B		&   $0.20255\pm0.00066$		&   $0.22623\pm0.00059$ 	& (16) \\
AD2615B			&   $0.212\pm0.012$		&   $0.233\pm0.013$ 		& (17) \\
KOI-126C		&   $0.2127\pm0.0026$		&   $0.2318\pm0.0013$ 		& (18) \\
CMDraA			&   $0.2130\pm0.0009$		&   $0.2534\pm0.0019$ 		& (19) \\
CMDraB			&   $0.2141\pm0.0010$		&   $0.2534\pm0.0019$ 		& (19) \\
HD24465B$^{\star}$			&   $0.233\pm0.002$		&   $0.244\pm0.001$             & This work \\
T-Lyr0-08070B		&   $0.24\pm0.019$		&   $0.265\pm0.010$ 		& (15) \\
SDSS-MEB-1B		&   $0.24\pm0.022$		&   $0.248\pm0.009$ 		& (20) \\
KOI-126B		&   $0.2413\pm0.0003$		&   $0.2543\pm0.0014$ 		& (18) \\
OGLE-TR-78B		&   $0.243\pm0.015$		&   $0.240\pm0.013$ 		& (21) \\
HATS551-027A		&   $0.244\pm0.003$		&   $0.261\pm0.009$ 		& (11) \\
AD2615A			&   $0.255\pm0.013$		&   $0.267\pm0.014$ 		& (17) \\
SAO106989B$^{\star}$		&   $0.256\pm0.005$		&   $0.326\pm0.012$		& This work \\
1RXSJ14727A		&   $0.2576\pm0.0085$		&   $0.2895\pm0.0068$ 		& (22) \\
1RXSJ14727B		&   $0.2585\pm0.0080$		&   $0.2895\pm0.0068$ 		& (22) \\
NSV-S6550671B		&   $0.260\pm0.02$		&   $0.290\pm0.01$ 		& (23) \\
SDSS-MEB-1A		&   $0.272\pm0.02$		&   $0.268\pm0.001$ 		& (20) \\
SDSSJ12120123		&   $0.273\pm0.002$		&   $0.306\pm0.007$ 		& (10) \\
LSPMJ1112B		&   $0.2745\pm0.0012$		&   $0.2978\pm0.005$ 		& (24) \\
GJ3236B			&   $0.281\pm0.015$		&   $0.3\pm0.015$ 		& (25) \\
GJ191			&   $0.281\pm0.014$		&   $0.291\pm0.025$ 		& (6) \\
HD213597B$^{\star}$		&   $0.286\pm0.012$ 		&   $0.344\pm0.01$  		& (26) \\
T-Boo0-0080		&   $0.315\pm0.01$		&   $0.325\pm0.005$ 		& (15) \\
LP133-373A		&   $0.34\pm0.02$		&   $0.330\pm0.014$ 		& (27) \\
LP133-373B		&   $0.34\pm0.02$		&   $0.330\pm0.014$ 		& (27) \\
T-cyg-1-01385		&   $0.345\pm0.034$		&   $0.360\pm0.019$ 		& (15) \\
WTS19e-3-08413B		&   $0.351\pm0.019$		&   $0.375\pm0.020$		& (28) \\
OGLE-TR-6		&   $0.359\pm0.025$ 		&   $0.393\pm0.018$ 		& (29) \\
TAur0-13378		&   $0.37\pm0.03$    		&   $0.37\pm0.02$ 		& (15) \\
GJ3236A			&   $0.376\pm0.016$		&   $0.3795\pm0.0084$ 		& (25) \\
WTS19c-3-01405B		&   $0.376\pm0.024$		&   $0.393\pm0.019$ 		& (28) \\
MG1-2056316B		&   $0.382\pm0.001$		&   $0.374\pm0.002$ 		& (30) \\
LSPMJ1112A		&   $0.3946\pm0.0023$		&   $0.3860\pm0.005$ 		& (24) \\
CuCnCB			&   $0.3980\pm0.0014$		&   $0.3908\pm0.0094$ 		& (31) \\
GJ411			&   $0.403\pm0.02$		&   $0.393\pm0.008$ 		& (6) \\
HD205403B$^{\star}$		&   $0.406\pm0.005$             &   $0.444\pm0.014$		& This work \\
WTS19c-3-01405A		&   $0.410\pm0.023$		&   $0.398\pm0.019$ 		& (28) \\
TCyg1-01385B		&   $0.43\pm0.02$		&   $0.40\pm0.02$ 		& (15) \\
CuCnCA			&   $0.4333\pm0.0017$		&   $0.4317\pm0.0052$ 		& (31) \\
MG1-646680B		&   $0.443\pm0.002$		&   $0.427\pm0.004$		& (30) \\
KELTJ041621-620046A	&   $0.447\pm0.05$		&   $0.540\pm0.032$ 		& (32) \\
WTS19e-3-08413A		&   $0.463\pm0.025$		&   $0.480\pm0.022$ 		& (28) \\
MG1-2056316A		&   $0.469\pm0.002$		&   $0.441\pm0.002$ 		& (30) \\
WTS19b-2-01387B		&   $0.481\pm0.017$		&   $0.479\pm0.013$		& (28) \\
MG1-78457B		&   $0.491\pm0.001$		&   $0.471\pm0.008$ 		& (30) \\
NSVS-01031772B		&   $0.498\pm0.0025$		&   $0.509\pm0.003$ 		& (33) \\
WTS19b-2-01387A		&   $0.498\pm0.019$		&   $0.496\pm0.013$ 		& (28) \\
MG1-646680A		&   $0.499\pm0.002$		&   $0.457\pm0.005$ 		& (30) \\
GJ887			&   $0.503\pm0.025$		&   $0.393\pm0.008$ 		& (6) \\
OGLE-TR-34		&   $0.509\pm0.038$		&   $0.435\pm0.033$ 		& (29) \\
UNSW2AB			&   $0.512\pm0.035$		&   $0.608\pm0.06 $ 		& (34) \\
T-Lyr-17236B		&   $0.523\pm0.006$		&   $0.525\pm0.052$ 		& (35) \\
MG1-78457A		&   $0.527\pm0.002$		&   $0.505\pm0.0075$ 		& (30) \\
MG1-116309B		&   $0.532\pm0.002$		&   $0.532\pm0.006$ 		& (30) \\
MG1-1819499B		&   $0.535\pm0.001$		&   $0.5\pm0.0085$ 		& (30) \\
HIP96515AaB		&   $0.54\pm0.03$		&   $0.55\pm0.03$ 		& (36) \\
NSVS-01031772A		&   $0.5428\pm0.0027$		&   $0.526\pm0.0028$ 		& (33) \\
MG1-506664B		&   $0.544\pm0.002$		&   $0.513\pm0.0055$ 		& (30) \\
NSVS-6550671A		&   $0.550\pm0.01$		&   $0.550\pm0.01$ 		& (23) \\
MG1-1819499A		&   $0.557\pm0.001$		&   $0.569\pm0.0022$ 		& (6) \\
MG1-116309A		&   $0.567\pm0.002$		&   $0.552\pm0.0085 $ 		& (30) \\
MG1-506664A		&   $0.584\pm0.002$		&   $0.560\pm0.0025$  		& (30) \\
BD-225866AaA		&   $0.5881\pm0.0029$		&   $0.614\pm0.045 $  		& (37) \\ 
BD-225866AaB		&   $0.5881\pm0.0029$		&   $0.598\pm0.045$ 		& (37) \\
HIP96515AaA		&   $0.59\pm0.03$		&   $0.64\pm0.01$ 		& (36) \\
V530OriB		&   $0.5955\pm0.0022$		&   $0.5873\pm0.0067$ 		& (38) \\
YYGemB			&   $0.5975\pm0.0047$		&   $0.6036\pm0.0057$ 		& (39) \\
EPIC211682657B$^{\star}$           &   $0.599\pm0.017$		&   $0.566\pm0.005$		& This work \\
UNSW2AA			&   $0.599\pm0.035$		&   $0.641\pm0.045$ 		& (34) \\
GuBooB			&   $0.600\pm0.006$		&   $0.624\pm0.016$ 		& (40) \\
YYGemA			&   $0.6009\pm0.0047$		&   $0.6196\pm0.0057$ 		& (39) \\
\enddata
\begin{tablenotes}
\item \scriptsize{References: (1) \cite{Triaud2013}; (2) \cite{Chaturvedi2016}; (3) \cite{Zhou2014}; \\
(4) \cite{Parsons2010}; (5)\cite{Parsons2012}; (6) \cite{Segransan2003}; (7) \cite{Beatty2007}; \\
(8) \cite{Ofir2012}; (9) \cite{Nefs2013}; (10) \cite{Pyrzas2012}; (11) \cite{Zhou2015}; \\
(12) \cite{Maxted2007}; (13) \cite{Gomez2014}; (14) \cite{Eigmuller2016}; (15)\cite{Fernandez2009};\\
(16) \cite{Doyle2011}; (17) \cite{Gillen2017}; (18) \cite{Carter2011}; (19) \cite{Morales2009}; \\
(20) \cite{Blake2008}; (21) \cite{Pont2005b}; (22) \cite{Hartman2011}; (23) \cite{Dimitrov2010}; \\
(24) \cite{Irwin2011}; (25) \cite{Irwin2009}; (26) \cite{Chaturvedi2014}; (27) \cite{Vaccaro2007}; \\
(28) \cite{Birkby2012}; (29) \cite{Bouchy2005}; (30) \cite{Kraus2011}; (31) \cite{Ribas2003}; \\
(32) \cite{Lubin2017}; (33) \cite{Lopez-Morales2006} (34) \cite{Young2006}; (35) \cite{Devor2008};\\
(36) \cite{Huelamo2009}; (37) \cite{Shkolnik2010}; (38) \cite{Torres2014}; (39) \cite{Torres2002}; \\
(40) \cite{Lopez-Morales2005} \\
$^{\star}$ -- PARAS spectra} \\
\end{tablenotes}
%\end{tiny}
%\vspace{-1.5cm}
\end{deluxetable*}

\software{{\tt{PARAS PIPELINE}} \citep{Chakraborty2014},  
		{\tt{PARAS SPEC}} \citep{Chaturvedi2016}, 
		 {\tt{ISOCHRONES}} \citep{Morton2015},  
		 {\tt{IRAF}} \citep{Tody1986,Tody1993},
		{\tt{REDUCE}} \citep{Piskunov2002},
		{\tt{PHOEBE}} \citep{Prsa2005,Prsa2016},
		{\tt{SPECTRUM}}\citep{Gray1999}}

%\bibliography{reference} % if your bibtex file is called example.bib
%\end{thebibliography}
\end{document}